\documentclass[12pt]{article}

\usepackage[top=80pt,bottom=85pt,left=67pt,right=67pt]{geometry}
\usepackage{amssymb,amsmath}
\usepackage{graphicx,color}
\usepackage{cancel}
\usepackage{float}
\usepackage{cite}
\usepackage[debug,pageanchor=false]{hyperref}
\definecolor{link}{rgb}{0,0,1}
\hypersetup{colorlinks=true,linkcolor=link,citecolor=link,urlcolor=link,linktocpage}

\makeatletter
\@addtoreset{equation}{section}
\makeatother

\renewcommand{\theequation}{\thesection.\arabic{equation}}

\newcommand{\beq}{\begin{equation}}
\newcommand{\eeq}{\end{equation}}
\newcommand{\bea}{\begin{eqnarray}}
\newcommand{\eea}{\end{eqnarray}}

\begin{document}
\begin{titlepage}

\begin{center}

\vskip .5in %.3in 
\noindent

{\Large \bf{AdS$_2$ geometries and non-Abelian T-duality in non-compact spaces}}

\bigskip\medskip

 Anayeli Ramirez\footnote{ramirezanayeli.uo@uniovi.es} \\

\bigskip\medskip
{\small 

Department of Physics, University of Oviedo,\\
Avda. Federico Garcia Lorca s/n, 33007 Oviedo, Spain}

\vskip 1.5cm 
\vskip .9cm %.6cm
     	{\bf Abstract }

\vskip .1in
\end{center}

\noindent
We obtain an AdS$_{2}$ solution to Type IIA supergravity with 4 Poincar\'e supersymmetries, via non-Abelian T-duality with respect to a freely acting SL(2,$\mathbf{R}$) isometry group, operating on the AdS$_3\times$S$^3\times$CY$_2$ solution to Type IIB. That is, non-Abelian T-duality on AdS$_3$. The dual background obtained fits in the class of AdS$_2\times$S$^3\times$CY$_2$ solutions to massive Type IIA constructed in \cite{Lozano:2020sae}. 
%We propose and study a quiver quantum mechanics dual to this solution coming from D0 and D4 instantons in the world-volumes of D8 and D4$'$ branes interacting with Wilson lines that we interpret as describing the baryon vertex of the D4-D8 brane intersection.  
We propose and study a quiver quantum mechanics dual to this solution that we interpret as describing the backreaction of the baryon vertex of a D4-D8 brane intersection.      

\noindent
 
\vfill
\eject

\end{titlepage}

\setcounter{footnote}{0}

\tableofcontents

\setcounter{footnote}{0}
\renewcommand{\theequation}{{\rm\thesection.\arabic{equation}}}

\section{Introduction}
Our understanding of four- and five-dimensional extremal black holes has extended our knowledge of supergravity backgrounds involving AdS$_2$ and AdS$_3$ geometries.    
For instance, an infinitely deep AdS$_2$ throat arises as the near horizon geometry of 4d extremal black holes that have associated an SL$(2,\mathbf{R})\times$U$(1)$ isometry, which 
%which has associated an SL$(2,\mathbf{R})\times$U$(1)$ isometry that 
includes the conformal group in 1d. Even if this limit is clear geometrically a microscopic understanding remains a demanding task \cite{Benini:2015eyy, Benini:2016rke, Cabo-Bizet:2017jsl}.
Via the AdS/CFT correspondence \cite{Maldacena:1997re} one might presume that there is a conformal quantum mechanics dual to these AdS$_2$ geometries. Nevertheless, AdS$_2$/CFT$_{1}$ pairs pose important conceptual puzzles \cite{Maldacena:1998uz, Denef:2007yt, Maldacena:2016hyu, Maldacena:2016upp} originated from the boundary of AdS$_2$ being non-connected \cite{Harlow:2018tqv}.
 
 Partial attempts at studying  AdS$_2$ and AdS$_3$ solutions in 10 and 11 dimensions, with vast and rich structures coming from the high dimensionality of the internal space, admitting many possible geometries, topologies and amounts of supersymmetry, have been carried out, \cite{Cvetic:2000cj, Gauntlett:2006ns,DHoker:2007mci,Chiodaroli:2009yw,Chiodaroli:2009xh,Kim:2013xza,Kelekci:2016uqv,Couzens:2017way,Dibitetto:2017klx,Corbino:2017tfl,Dibitetto:2018iar,Dibitetto:2018gtk,Corbino:2018fwb,Macpherson:2018mif,Hong:2019wyi,Lozano:2019emq,Lozano:2019jza,Lozano:2019zvg,Lozano:2019ywa,Legramandi:2019xqd,Dibitetto:2019nyz,Lust:2020npd,Corbino:2020lzq,Lozano:2020bxo,Chen:2020mtv,Faedo:2020lyw,Faedo:2020nol,Lozano:2020txg,Lozano:2020sae,Passias:2020ubv,Legramandi:2020txf,Lozano:2021rmk,Balaguer:2021wjf,Zacarias:2021pfz}. 
 In particular, recent progress has been reported on the construction of new AdS$_3$ solutions with four Poincar\'e supersymmetries \cite{Couzens:2017way,Macpherson:2018mif,Lozano:2019emq,Lozano:2020bxo,Faedo:2020nol} as well as on the identification of their 2d (half-maximal BPS) dual CFTs \cite{Lozano:2019jza, Lozano:2019zvg, Lozano:2019ywa,Lozano:2020bxo,Faedo:2020nol,Roychowdhury:2021eas}. %, by providing AdS$_3$/CFT$_2$ pairs where a precise family of quivers that flow to (0,4) fixed points at low energies was proposed.
 %In particular, recent progress has been reported for 2d CFT half-maximal BPS providing  AdS$_3$/CFT$_2$ pairs (see \cite{Lozano:2019jza, Lozano:2019zvg, Lozano:2019ywa}), where a precise family of quivers that flow to (0,4) fixed points at low energies was proposed. 
 In the same vein, AdS$_2$/CFT$_1$ pairs have been explored as a natural extension of AdS$_3$/CFT$_2$ pairs through T-duality \cite{Lozano:2020txg} and double analytical continuation, \cite{Lozano:2020sae,Lozano:2021rmk}, in each case, providing different families of quiver quantum mechanics with four Poincar\'e supersymmetries.

Part of the motivation for this work is to construct AdS$_2$ solutions through non-Abelian T-duality acting on AdS$_3$ spaces. Non-Abelian T-duality (NATD) was introduced in the 90\'{}s \cite{delaOssa:1992vci} as a transformation of the string $\sigma$-model, generalising to non-Abelian isometry groups the path integral approach to Abelian T-duality put forward in \cite{Rocek:1991ps}. From these studies other important groundwork arose, see for example \cite{Giveon:1993ai, Alvarez:1994np, Sfetsos:1994vz,Lozano:1995jx,Sfetsos:1996pm}.  
In spite of this initial progress and unlike its Abelian counterpart, the NATD transformation did not reach the status of a string theory symmetry \cite{Alvarez:1993qi,Alvarez:1994np,Alvarez:1994wj,Elitzur:1994ri,Klimcik:1995ux, Lozano:1995jx, Sfetsos:1996pm}, due to two main difficulties. Firstly, NATD has only been worked out as a transformation in the worldsheet for spherical topologies (namely, at tree level in string perturbation theory) and second, the conformal symmetry of the string $\sigma$-model is only known to survive the NATD transformation at first order in $\alpha'$.
%NATD is only known to respect conformal symmetry to first order in $\alpha'$

%The NATD transformation was carried out between worldsheet field theories and its status as full symmetry of string perturbation theory remains a technically hard problem \cite{Alvarez:1993qi,Alvarez:1994np, Alvarez:1994wj,Elitzur:1994ri,Klimcik:1995ux, Lozano:1995jx, Sfetsos:1996pm}, unlike its Abelian counterpart does.
    
  Sfetsos and Thompson \cite{Sfetsos:2010uq} reignited the interest in NATD by showing that it can be successfully used as a solution generating technique in supergravity, with the derivation of the transformation rules of the RR sector. This study was initiated with the dualisation of the AdS$_5\times$S$^5$ and AdS$_3\times$S$^3\times$CY$_2$ backgrounds  with respect to a freely acting SU(2) isometry group (SU(2)-NATD). This work was of particular interest to tackle the r\^ole NATD might have in the context of AdS/CFT correspondence. 
  In this vein, interesting examples of AdS spacetimes generated through NATD in different contexts have been constructed to date \cite{Sfetsos:2010uq,Lozano:2011kb,Lozano:2012au,Itsios:2012zv,Itsios:2012dc,Itsios:2013wd,Barranco:2013fza,Macpherson:2013zba,Lozano:2013oma,Zacarias:2014wta,Lozano:2014ata,Sfetsos:2014tza,Kelekci:2014ima,Macpherson:2014eza,Kooner:2014cqa,Lozano:2015cra,Macpherson:2015tka,vanGorsel:2017goj}.
Holographically, the field theoretical interpretation of NATD was first addressed in  \cite{Lozano:2016kum,Lozano:2016wrs,Lozano:2017ole,Itsios:2017cew,Lozano:2019ywa}, where the main conclusion is that NATD changes the field theory dual to  the original theory. %\textcolor{blue}{This fact is ascribed that NATD has not been proven to be a string theory symmetry}. 
  Remarkably, in all examples so far of NATD in supergravity --in the context of holography-- the dualisation took place with respect to a freely acting SU(2) subgroup of the entire symmetry group of the solutions. 
  
The main purpose of this work is to construct an AdS$_2$ solution to massive Type IIA supergravity acting with NATD on the well-known D1-D5 near horizon system. Here the dualisation is performed with respect to a freely acting SL(2,$\mathbf{R}$) group (SL(2,$\mathbf{R}$)-NATD). Second, we give a proposal for its dual superconformal quantum mechanics, in terms of D0 and D4 colour branes coupled to D4$'$ and D8 flavour branes, inspired by the results in \cite{Lozano:2020sae}.

    The organisation of the paper is as follows. In section \ref{NATDSection}, we develop the technology necessary to construct solutions through SL(2,$\mathbf{R}$)-NATD. In the same section we apply these results to the D1-D5 near horizon system, generating a new AdS$_2\times$S$^3\times$CY$_2$ geometry foliated over an interval. The brane set-up, charges and holographic central charge are carefully studied. Section \ref{typeCreview} contains a  summary of the infinite family of AdS$_2$ solutions to massive Type IIA supergravity with four Poincar\'e supersymmetries constructed in  \cite{Lozano:2020sae}, as well as of the quiver quantum mechanics proposed there as duals to these geometries. In section \ref{QM-NATD}, we show that our SL(2,$\mathbf{R}$)-NATD solution provides an explicit example in the classification in \cite{Lozano:2020sae}. %The solutions generated by SU(2)-NATD are generically non-compact manifolds, as we will see the SL(2,$\mathbf{R}$)-NATD solution generating technique preserves this non-compactness, indicating that a completion of the background is needed to provide a holographic interpretation of the newly AdS solution.
%\textcolor{blue}{The solutions generated by SU(2)-NATD are generically non-compact manifolds even if the group used in the dualisation procedure is compact, this fact is due to the new generated variables now live in the Lie algebra of the Lie group involved in the dualisation. %, eh non-compact by its construction 
%As we will see the SL(2,$\mathbf{R}$)-NATD solution generating technique inherits this non-compactness, suggesting a completion of the background in order to provide an holographic interpretation of the newly AdS solution.}
  %As in the SU(2)-NATD case, the NATD generating technique under a non-compact group (such as SL(2,$\mathbf{R}$)) generates a non-compact internal geometry, which is reflected in a long dual quiver. 
 At the end of this section we study an explicit completion of this solution and propose a quiver quantum mechanics that admits a description in terms of interactions between Wilson lines and  D0 and D4 instantons in the world-volumes of the D4$'$ and D8 branes. 
 Our conclusions are contained in Section \ref{conclusions}.                
%%%%%%%%%%%%%%%%%%%%%%%%%%%%%%%%%
\section{NATD of AdS$_3\times$S$^3\times$CY$_2$ with respect to a freely acting SL$(2,\mathbf{R})$}\label{NATDSection}

In this section we review the dualisation procedure and apply it to the AdS$_3\times$S$^3\times$CY$_2$ solution of Type IIB supergravity. We address the construction of the brane set-up, Page charges and holographic central charge of the resulting background and propose a quiver quantum mechanics that flows in the IR to the superconformal quantum mechanics dual to our solution.

%%%%%%%%%%%%%%%%  
\subsection{NATD with respect to SL$(2,\mathbf{R})_L$}

The study of NATD as a solution generating technique in supergravity was initiated in \cite {Sfetsos:2010uq}, where the dualisation was carried out with respect to a freely acting SU(2) isometry group. Since then, several works have taken advantage of this technology to generate new AdS solutions, some of which avoiding previously existing classifications (see for instance \cite{Lozano:2012au,Lozano:2015bra,Macpherson:2014eza,Bea:2015fja}). 
Most of these examples possess rich isometry groups containing at least an SU(2) factor that can be used to dualise. Instead, in this work we will use a non-compact, freely acting, SL(2,$\mathbf{R}$) group to dualise. This is the first time that NATD with respect to a non-compact isometry group has been applied as a solution generating technique in supergravity\footnote{In \cite{Lozano:2021rmk},  SL$(2,\mathbf{R})$-NATD was used to find an explicit example --with brane sources-- in the class of AdS$_2\times$S$^2\times$CY$_2$ solutions fibered over a 2d Riemann surface constructed in \cite{Chiodaroli:2009yw}.}. 
 Following \cite{Alvarez:1993qi} we perform the NATD transformation with respect to one of the freely acting SL$(2,\mathbf{R})$ isometry groups of the AdS$_3$ subspace of the AdS$_3\times$S$^3\times$CY$_2$ solution of Type IIB supergravity. We start reviewing the necessary technology.

Consider a bosonic string $\sigma$-model that supports an SL$(2,\mathbf{R})$ isometry, such that the NS-NS fields can be written as,
\begin{gather}
	\label{ssm}
	\begin{split}
		&ds^2=\frac{1}{4}g_{\mu\nu}(x)L^\mu L^\nu+G_{i\mu}(x)dx^iL^\mu+G_{ij}(x)dx^idx^j,\\
		B_2=\frac{1}{8}b_{\mu\nu}(x)&L^\mu\wedge L^\nu +\frac{1}{2}B_{i\mu}(x)dx^i\wedge L^\mu+
		B_{ij}(x)dx^i\wedge dx^j, \qquad \Phi=\Phi(x),
	\end{split}
\end{gather}
where $x^i$ are the coordinates in the internal manifold, for $i,j=1,2,...,7$, and $L^{\mu}$ are the SL$(2,\mathbf{R})$ left-invariant Maurer-Cartan forms,
\begin{gather}
	\label{MCforms}
	L^\mu=-i\textrm{Tr}(t^\mu g^{-1}\text{d}g),\qquad \textrm{which obey},\quad \textrm{d}L^\mu=\frac{1}{2}{f^\mu}_{\alpha\nu}L^\alpha\wedge L^\nu,
\end{gather}
where ${f^\mu}_{\alpha\nu}$ are the structure constants of SL(2,$\mathbf{R}$). % In turn, the $\sigma$-model in invariant under $g\to\lambda g$ for $\lambda\in$ SL(2,$\mathbf{R}$). 
 The generators of the $sl(2,\mathbf{R})$ algebra can be obtained by analytically continuing the $su(2)$ generators as,
\begin{gather}
	t^a=\frac{\tau_a}{\sqrt{2}},\qquad%\nonumber\\ &&
	\textrm{with}\quad
	\tau_1=\left(\begin{matrix}
		&0 & i\\
		&i& 0
	\end{matrix}\;\;\right),\quad \tau_2=\left(\begin{matrix}
		&0 & -i\\
		&i& 0
	\end{matrix}\;\;\right),	\quad \tau_3=\left(\begin{matrix}
		&i & 0\\
		&0& -i
	\end{matrix}\;\;\right).		
\end{gather}
These generators satisfy\footnote{We take $g^{\mu\nu}=-\textrm{Tr}(t^\mu t^\nu)$ in order to have $(+,-,+)$ signature.},
\begin{gather}
	\text{Tr}(t^at^b)=(-1)^a\delta^{ab},\quad\quad\quad	[t^1,t^2]=i\sqrt{2}t^3,\quad	[t^2,t^3]=i\sqrt{2}t^1,\quad	[t^3,t^1]=-i\sqrt{2}t^2.
\end{gather}

The group element $g\in$ SL(2,$\mathbf{R}$) depends on the target space isometry directions, realising an SL(2,$\mathbf{R}$) group manifold. Here the group manifold is an AdS$_3$ space.
The geometry described by \eqref{ssm} is then manifestly invariant under  $g\to\lambda^{-1} g$ for $\lambda\in$ SL(2,$\mathbf{R}$).
 We parametrise an SL(2,$\mathbf{R}$) group element in the following fashion,
\begin{gather}
	g=e^{\frac{i}{2}t\tau_3}e^{\frac{i}{2}\theta\tau_2}e^{\frac{i}{2}\eta\tau_3},	
	\qquad\text{with} \qquad 0\leq\theta\leq\pi,\;\; 0\leq t< \infty,\;\; 0\leq\eta< \infty,
\end{gather}
%here we have used a parametrisation similar to the Euler parametrisation for SU(2).
which is closely related to the Euler angles parametrising SU(2).
Thus, the left-invariant forms \eqref{MCforms} are given by,
\begin{gather}
	\begin{split}
		\label{Maurer}
		L^1=\sinh{\eta}\text{d}\theta-\cosh{\eta}&\sin{\theta}\text{d}t,\qquad
		L^2=\cosh{\eta}\text{d}\theta-\sinh{\eta}\sin{\theta}\text{d}t,\\
		&L^3=-\cos{\theta}\text{d}t-\text{d}\eta.
	\end{split}
\end{gather}
%all the coordinate dependence on the SL$(2,\mathbf{R})$ is contained in the \eqref{Maurer} forms, the rest of information is included on the spectator fields.

A string propagating in the geometry given by \eqref{ssm} is described by the non-linear $\sigma$-model,
\begin{gather}
	\label{sm}
	S=\int \textrm{d}\sigma^2\left(E_{\mu\nu}L_{+}^{\mu}L_{-}^{\nu}+Q_{i\mu}\partial_+x^iL_-^\mu+Q_{\mu i}L_+^\mu\partial_-x^i+Q_{ij}\partial_+x^i\partial_-x^j\right),\\
	\textrm{with}\qquad E_{\mu\nu}=g_{\mu\nu}+b_{\mu\nu},\qquad Q_{i\mu}=G_{i\mu}+B_{i\mu},\qquad Q_{\mu i}=G_{\mu i}+B_{\mu i},\qquad Q_{ij}=G_{ij}+B_{ij}.\nonumber
\end{gather}
and $L_{\pm}^\mu$ are the left-invariant forms pulled back to the worldsheet. This $\sigma$-model is also invariant under $g\to\lambda^{-1} g$ for $\lambda\in$ SL(2,$\mathbf{R}$).

The SL(2,$\mathbf{R}$) non-Abelian T-dual solution for the $\sigma$-model \eqref{sm} is constructed as in \cite{delaOssa:1992vci}, introducing covariant derivatives, $\partial_{\pm}g\to D_{\pm}g=\partial_{\pm}g-A_{\pm}g$, in the Maurer-Cartan forms but enforcing the condition that the gauge field is non-dynamical with the addition to the action of a Lagrange multiplier term,
\begin{gather}
 -i\textrm{Tr}(vF_{\pm}),%\qquad \text{where}\qquad F_{\pm}=\partial_+A_--\partial_-A_+-[A_+,A_-].
\end{gather}
where $F_{\pm}=\partial_+A_--\partial_-A_+-[A_+,A_-]$ is the field strength for the gauged fields $A_{\pm}$. $v$ is a vector that takes values in the Lie algebra of the SL(2,$\mathbf{R}$) group and it is coupled to the field strength, $F_{\pm}$.
%The \textcolor{red}{SL(2,$\mathbf{R}$) non-Abelian T-dual solution to \eqref{sm}} is constructed, as in \cite{delaOssa:1992vci}, introducing covariant derivatives, $\partial_{\pm}g\to D_{\pm}g=\partial_{\pm}g-A_{\pm}g$, in the Maurer-Cartan forms, together with a Lagrange multiplier term, $-i\textrm{Tr}(vF_{\pm})$, enforcing the condition that the gauge field is non-dynamical. Where $F_{\pm}=\partial_+A_--\partial_-A_+-[A_+,A_-]$ is the field strength for the gauged fields, $A_{\pm}$, and $v$ is a vector that lives in Lie algebra of the Lie group involved in the dualisation.
In this way, the total action is invariant under,
\begin{gather}
	g\to\lambda^{-1}g,\quad A_{\pm}\to \lambda^{-1}(A_{\pm}\lambda-\partial_{\pm}\lambda),\quad v\to\lambda^{-1}v\lambda,\quad\text{with}\quad\lambda(\sigma^+,\sigma^-)\in \text{SL}(2,\mathbf{R}). 
\end{gather}
After integrating out the Lagrange multiplier and fixing the gauge, we recover the original non-linear $\sigma$-model.
%Notice that the original non-linear $\sigma$-model is recovered upon fixing the pure gauge fields to zero. 
On the other hand, by integrating by parts the Lagrange multiplier term one can solve for the gauge fields and obtain the dual $\sigma$-model, that still relies on the  parameters $t,\theta,\eta$ and the Lagrange multipliers. In order to preserve the number of degrees of freedom, the redundancy is fixed by choosing a gauge fixing condition, for instance $g=\mathbb{I}$, which implies $t=\theta=\eta=0$.
The resulting action reads,
\begin{gather}
	\label{dualsm}
\hat{S}=\int \textrm{d}\sigma^2\left[Q_{ij}\partial_+x^i\partial_-x^j+(\partial_+v_\mu+\partial_+x^iQ_{i \mu})M_{\mu\nu}^{-1}(\partial_-v_\nu-Q_{\nu i}\partial_+x^i)\right],\\
	\textrm{with}\qquad M_{\mu\nu}=E_{\mu\nu}+{f^\alpha}_{\mu\nu}v_{\alpha}.\nonumber
\end{gather}
%from which one can obtain the NS-NS fields of the dual background. 
In this action the parameters $t,\theta,\eta$ have been replaced by the Lagrange multipliers $v_i$, $i=1,2,3$, which live in the Lie algebra of SL(2,$\mathbf{R}$), %. Namely, the group manifold coordinates of SL(2,$\mathbf{R}$) --the Euler angles-- to get mapped to elements on the Lie algebra of the uncompact $sl(2,\mathbf{R})$, which 
this is non-compact, by its construction as a vector space.% \textcolor{blue}{The non-compactness will be reflected at level of the metric --and after an adequate parametrisation of the Lagrange multipliers-- by the substitution of the AdS$_3$ space of the original background to an AdS$_2\times\mathbf{R}^+$ space, where in addition of the AdS$_2$ factor a non-compact radial direction is generated in the dual geometry.}
 %when the AdS$_3$ space of the original background is substituted   
 
In particular, the solutions generated by SU(2)-NATD are non-compact manifolds  
 	even if the group used in the dualisation procedure is compact, this is because the new variables live in the Lie algebra of the dualisation group.
 	%this is due to the fact the new variables now live in the Lie algebra of the Lie group involved in the dualisation. 
 	As we see, the SL(2,$\mathbf{R}$)-NATD solution generating technique inherits this non-compactness. At the level of the metric and using the following parametrisation for the Lagrange multipliers,
 \begin{gather}
 		\label{likepolarcoordinades}
 		v=(\rho\cos\tau\cosh\xi,\rho\sinh\xi,\rho\sin\tau\cosh\xi),
 \end{gather}
 	the original AdS$_3$ space is replaced by an AdS$_2\times\mathbf{R}^+$ space, where besides the AdS$_2$ factor (in which the remaining SL(2,$\mathbf{R}$) symmetry is reflected) a non-compact radial direction is generated in the internal space.
 
Furthermore, from the path integral derivation the dilaton receives a 1-loop shift, leading to a non-trivial dilaton in the dual theory, given by, 
\begin{eqnarray}
	\hat{\Phi}(x,v)=\Phi(x)-\frac{1}{2}\log(\det M).
\end{eqnarray}  
A similar shift in the dilaton was obtained in Abelian T-duality \cite{delaOssa:1992vci}, in such a case $M$ is the metric component in the direction where the dualisation is carried out.   

%\textcolor{blue}{this deviation in the dilaton is the suitable shift to maintain the conformal invariance of the dual solution \cite{delaOssa:1992vci}}.   

The transformation rules for the RR fields %was the essential new derivation 
was the new input in \cite{Sfetsos:2010uq} which allowed to use NATD as a solution generating technique in supergravity. This was done using a spinor representation approach. 
The derivation relied on the fact that left and right movers transform differently under NATD, and therefore lead to two different sets of frame fields for the dual geometry. In the SL(2,$\mathbf{R}$)-NATD case, we also have two different sets of frame fields, %Both frame fields 
which define the same dual metric obtained from \eqref{dualsm}, and must therefore be related by a Lorentz transformation, ${\Lambda^\alpha}_\beta$. In turn,  this Lorentz transformation acts on spinors through a  matrix $\Omega$, defined by the invariance property of gamma matrices, 
\begin{gather}
\Omega^{-1}\Gamma^\alpha\Omega={\Lambda^\alpha}_\beta\Gamma^\beta,	
	\end{gather}
%which can be derived from the Lorentz transformation relating the vielbeins for left and right movers of the dual $\sigma$-model \eqref{dualsm}.  
%The idea is based on the construction of a Lorentz transformation matrix $\Lambda$ relating two different sets of frame fields for the metric of the dual $\sigma$-model \eqref{dualsm}. This Lorentz transformation acts on the spinors through a  matrix $\Omega$, $\Omega^{-1}\Gamma^\alpha\Omega={\Lambda^\alpha}_\beta\Gamma^\beta$.
and given that the RR fluxes can be combined to form bispinors,
\begin{gather}
	\begin{split}
		P=&\frac{e^{\Phi}}{2}\sum_{n=0}^{4}\cancel{F}_{2n+1},\qquad \hat{P}=\frac{e^{\hat{\Phi}}}{2}\sum_{n=0}^{5}\cancel{\hat{F}}_{2n},\qquad\text{with}\quad \cancel{F}_p=\frac{1}{p!}\Gamma_{\nu_1...m_p}F_{p}^{\nu_1\nu_2...m_p},
	\end{split}
\end{gather}
  one can finally extract their transformation rules by right multiplication with the $\Omega^{-1}$ matrix on the RR bispinors, 
\begin{gather}\label{ActionRRsector}
\hat{P}=P\cdot\Omega^{-1},	
	\end{gather}
where $\hat{P}$ are the dual RR bispinors. %\footnote{The matrix $\Omega$ can be obtained by analytically continued the matrix $\Omega$ of SU(2)-NATD.}.
 Notice that the action \eqref{ActionRRsector} on the RR sector is from a Type IIB to a IIA solution. If starting from a Type IIA to a IIB solution instead, the r\^ole of $P$ and $\hat{P}$ is swapped. The knowledge of the transformation rules for the RR sector guarantees that starting with a solution to Type II supergravity the dual background is also a solution. 

The technology reviewed in this section allows us to consider a non-compact space like AdS$_3$, which posses an SO$(2,2)\cong\textrm{SL}(2,\mathbf{R})_L\times\textrm{SL}(2,\mathbf{R})_R$ isometry group. After performing the dualisation with respect to a freely acting $\textrm{SL}(2,\mathbf{R})$ group the isometry gets reduced to just $\textrm{SL}(2,\mathbf{R})$, which is geometrically realised by an AdS$_2$ factor in the dual geometry. 
 Further, as we explained before, the dual geometry acquires a non-compact direction, that now belongs to the internal space.  %and no group manifold can be assigned to them.    
% Further, as usual in NATD, the dual geometry acquires a non-compact direction at internal space. 

 In the next section we will apply this technology to the AdS$_3\times$S$^3\times$CY$_2$ geometry to produce an AdS$_2\times$S$^3\times$CY$_2\times$I solution in massive Type IIA supergravity, which fits in the classification in \cite{Lozano:2020sae}.   

%%%%%%%%%%%%%%
\subsection{Dualisation of the AdS$_3\times$S$^3\times$CY$_2$ background} 

We consider IIB string theory on $\mathbf{R}^{1,1}\times\mathbf{R}^4\times$CY$_2$ where we include $Q_{\text{D}1}$ D1-branes and $Q_{\text{D}5}$ D5-branes as is shown in the brane set-up depicted in Table \ref{branesetupD1D5}.
%M$_4$, where M$_4$ is an internal compact manifold which may be taken to be a CY$_2$. 
%The brane setup for this system is shown in the Table \ref{branesetupD1D5}. 

\begin{table}[ht]
	\begin{center}
		\begin{tabular}{| l | c | c | c | c| c | c| c | c | c | c|}
			\hline		    
			& $x^0$ & $x^1$ & $x^2$ & $x^3$ & $x^4$ & $x^5$ & $x^6$ & $x^7$ & $x^8$ & $x^9$  \\ \hline
			D1 & x &  & & & &x& & & & \\ \hline
			D5 & x&x&x&x&x&x& & & & \\ \hline
		\end{tabular} 
	\end{center}
	\caption{The set-up for $Q_{\text{D}1}$ D1-branes wrapped on $x^5$ and $Q_{\text{D}5}$ D5-branes wrapped on $x^5$ and CY$_2$. This system preserves (4,4) supersymmetry. The field theory lives in the $x^0$ and $x^5$ directions and $x^1,x^2,x^3,x^4$ parameterise the CY$_2$. The SO(4) R-symmetry is geometrically realised in the $x^6,x^7,x^8,x^9$ directions.}  
	\label{branesetupD1D5}
\end{table} 

The AdS$_3\times$S$^3\times$CY$_2$ background arising in the near horizon limit of the D1-D5 system depicted in Table \ref{branesetupD1D5} is,
\begin{gather}
\begin{split}
\label{AdS3S3T4}
&ds_{10}^2=4L^2 ds^2_{\text{AdS}_3}+M^2 ds^2_{\text{CY}_2}+4L^2 ds^2_{\text{S}^3},\qquad\qquad
e^{2\Phi}=1,\\
F_3=8L^2&(\text{vol}_{\text{S}^3}+\text{vol}_{\text{AdS}_3}),\qquad\qquad
F_7=-8L^2M^4(\text{vol}_{\text{S}^3}+\text{vol}_{\text{AdS}_3})\wedge\text{vol}_{\text{CY}_2}\, .
\end{split}
\end{gather}
Here we will use Vol$_{\text{CY}_2}=(2\pi)^4$.
%The corresponding D1 and D5 brane charges are given by,
%\begin{gather}
%Q_{\text{D}1}=\frac{1}{(2\pi)^6}\int_{\text{S}^3\times \text{T}^4}F_7=4L^2M^4,\qquad
%Q_{\text{D}5}=\frac{1}{(2\pi)^2}\int_{\text{S}^3} F_3=4L^2.
%\end{gather}

Following the rules explained in the previous section, the SL$(2,\mathbf{R})$-NATD transformation of the background \eqref{AdS3S3T4} gives rise to the following geometry,
\begin{gather}
\begin{split}
\label{NATD1}
ds_{10}^2&= \frac{L^2\rho^2}{\rho^2-4L^4}ds^2_{\text{AdS}_2}+4L^2ds^2_{\text{S}^3}+M^2 ds^2_{\text{CY}_2}+\frac{d\rho^2}{4L^2}, \\
e^{2\Phi}&=\frac{4}{L^2(\rho^2-4L^4)},\qquad\qquad
B_2=-\frac{\rho^3}{2(\rho^2-4L^4)}\text{vol}_{\text{AdS}_2},\\
F_0&=L^2,\qquad
F_2=-\frac{L^2\rho^3}{2(\rho^2-4L^4)}\text{vol}_{\text{AdS}_2},\qquad
F_4=-L^2 (M^4 \text{vol}_{\text{CY}_2}-2\rho\;\text{d}\rho\wedge\text{vol}_{\text{S}^3}),\\
F_6&= \frac{L^2 \rho^2}{2(\rho^2-4L^4)}(\rho M^4\text{vol}_{\text{CY}_2}-8L^4\text{d}\rho\wedge\text{vol}_{\text{S}^3})\wedge \text{vol}_{\text{AdS}_2},\\
F_8&=2L^2M^4\rho\;\text{vol}_{\text{S}^3}\wedge \text{vol}_{\text{CY}_2} \wedge\text{d}\rho,\quad
F_{10}=-\frac{4L^6 \rho^2M^4}{\rho^2-4L^4}\text{vol}_{\text{AdS}_2}\wedge\text{vol}_{\text{S}^3}\wedge \text{vol}_{\text{CY}_2}\wedge\text{d}\rho.
\end{split}
\end{gather}
Here we have parametrised  the Lagrange multipliers as in  \eqref{likepolarcoordinades} in order to manifestly realise the  SL(2,$\mathbf{R}$) residual global symmetries. Indeed, from the original SO$(2,2)$ isometry group, after the dualisation, an SL(2,$\mathbf{R}$) subgroup survives, which is geometrically realised by a warped AdS$_2\times\mathbf{R}^+$ subspace.       

The background \eqref{NATD1} is a solution to the massive Type IIA supergravity EOMs.  As we will see in Section \ref{QM-NATD}, it is an explicit solution in the classification provided in \cite{Lozano:2020sae}. In order to have the right signature and avoid singularities we are forced to set $\rho^2-4L^4>0$. Namely, we get a well-defined geometry for, 
\begin{gather}
\rho>\rho_0=2L^2,\label{r0natd}
\end{gather}
where the $\rho$ coordinate begins.  
%In addition,
%\begin{eqnarray}
%H_7=8L^2M^4\frac{(\rho^2-12L^4)}{(\rho^2-4L^4)}\text{vol}_{\text{S}^3}	\wedge \text{vol}_{\text{T}^4}
%\end{eqnarray}

The asymptotic behaviour of the metric and dilaton in \eqref{NATD1} at the beginning of the space, around $\rho=\rho_0$ is,
\begin{equation}
\label{F1-singularity}
ds^2\sim \frac{a_1}{x}ds^2_{\text{AdS}_2}+a_2	ds^2_{\text{S}^3}+M^2ds^2_{\text{CY}_2}+a_3 dx^2,\qquad\qquad e^{\Phi}\sim a_4x^{-1/2},
\end{equation}
with  $x=\rho-\rho_0>0$ and $a_i$ are constants. Here the warp factor reproduces the behaviour of an OF1 plane  extended in AdS$_2$ and smeared over S$^3$, this is also consistent with additional coincident fundamental strings if they are smeared on the S$^3$ and the CY$_2$.
% This singularity is recognised as associated to a fundamental string with worldsheet in AdS$_2$ and smeared over the S$^3$ and the CY$_2$.
%with  $x=\rho-2L^2$ at the beginning  and  $x=2L^2P-\rho$ at the end in the $\rho$ coordinate. Here $a_i$ are constants. This singularity is recognised as the singularity of a fundamental string smeared over S$^3\times$CY$_2$.
%\textcolor{blue}{The fact that the Lagrange multipliers take values in the Lie algebra suggest that global information of the dual background cannot be read into from the transformation itself. Thus,  we consider that internal geometry is globally unknown but compact. In this vein, a completion is needed for our dual background. We will tackle this point in Section \ref{field-theory} but advance that at both ends of the space the configuration is identified as a smeared fundamental string given by the expression \eqref{F1-singularity}.}      
%Due to the Lagrange multipliers live in the Lie algebra of the SL(2,$\mathbf{R}$) group, the global information of the background cannot be inferred from the transformation itself. In this way, a completion for the solution given in \eqref{NATD1} is needed. In Section \ref{field-theory}, we will provide a concrete completion, where at both ends of the space a smeared fundamental string,  like \eqref{F1-singularity}, is identified.
%As we mentioned before, the global information of the background cannot be inferred from the transformation itself, due to the Lagrange multipliers take values in the Lie algebra, $\mathbf{R}$.
Further, in Section \ref{field-theory}, we will provide a concrete completion for the background \eqref{NATD1}, where at both ends of the space the behaviour given in \eqref{F1-singularity} is identified.

We conclude this section with some comments about the supersymmetry of the solution  \eqref{NATD1}. On one hand, 
as we mentioned before (and we will show in Section \ref{QM-NATD}) the background \eqref{NATD1} fits in the class of AdS$_2\times$S$^3\times$CY$_2\times$I solutions to massive Type IIA  constructed in  \cite{Lozano:2020sae}, which contain eight supersymmetries, four Poincar\'e and four conformal.
 Second, it is well established by now \cite{Sfetsos:2010uq,Itsios:2012dc} that performing non-Abelian T-duality on a round 3-sphere projects out the spinors charged under either the SU(2)$_L$ or SU(2)$_R$ subgroup of the global SO(4) factor of S$^3$, leaving the rest intact. This amounts to a halving of supersymmetry in the non-Abelian T-dual of AdS$_3\times$S$^3\times$CY$_2$ \cite{Sfetsos:2010uq}. The SL(2,$\mathbf{R}$)-NATD works analogously, this time one projects out the spinors charged under one of the SL(2,$\mathbf{R}$) factors of the global SO$(2,2)\cong\textrm{SL}(2,\mathbf{R})_L\times\textrm{SL}(2,\mathbf{R})_R$ isometry, keeping the rest intact. As such SL(2,$\mathbf{R}$)-NATD on the AdS$_3\times$S$^3\times$CY$_2$ solution also reduces the supersymmetry by half. That this mirrors the halving of the supersymmetries as in the SU(2)-NATD case is hardly surprising, the solutions are after all related by a double analytic continuation (as we will explain around the equation \eqref{diagramaeq}).

\subsection{Brane set-up and charges}\label{set-upNATD}

Non-Abelian T-dualisation under a freely acting SU(2) subgroup of an $\text{SO}(4)$ symmetry reduces the isometry group to $\text{SU}(2)$. 
Geometrically, the S$^3$ is replaced by its Lie algebra, $\mathbf{R}^3$, which is locally $\mathbf{R}\times$S$^2$. This isometry is reflected in the dual fields, %namely we can guarantee the existence of non-trivial 2-cycles in the geometry. In particular 
for instance a $B_2$ over the S$^2$ is generated after the dualisation, which is $\rho$ dependent (like the $B_2$ in \eqref{NATD1}). This $\rho$ dependence in $B_2$ implies that large gauge transformations must be included such that $\frac{1}{4\pi^2}|\int B_2|$ remains in the fundamental region as we move in the $\rho$ direction. This argument was developed in \cite{Lozano:2013oma, Lozano:2014ata, Lozano:2016kum} where the non-compactness in the $\rho$ coordinate --in backgrounds like  \eqref{NATD1}-- was addressed with the introduction of large gauge transformations in the dual geometry.

%\textcolor{blue}{For instance, the SU(2)-NATD technology produces an antisymmetric Kalb-Ramond tensor over the non-trivial S$^2$ which is bounded in order to determine the range of the $\rho$ coordinate. This argument was developed in \cite{Lozano:2013oma, Lozano:2014ata, Lozano:2016kum}, where the authors showed that non-compactness in the coordinate $\rho$ in backgrounds like one of the expression \eqref{NATD1} is reflected in the existence of large gauge transformations.}

%%%%%%%%%%%%%%%%%%%%%%%%%%%%%%%%%%%%%%%
%In usual cases, the NATD with respect to a freely acting SU(2) subgroup of SO(4) R-symmetry group becomes the S$^3$ to locally $\mathbf{R}\times$S$^2$, in that case the NATD technology produces an antisymmetric Kalb-Ramond tensor over the non-trivial S$^2$ which is bounded in order to determine the range of the $\rho$ coordinate, see \cite{Lozano:2013oma, Lozano:2014ata, Lozano:2016kum}. In addition, such $B_2$ implies the presence of NS5 branes in the geometry. 
%In the background \eqref{NATD1}, 
The SL$(2,\mathbf{R})$-NATD, as shown in the previous section,  %respect to a freely acting  group 
produces an antisymmetric Kalb-Ramond tensor over the AdS$_2$ directions, signaling the presence of fundamental strings in the solution. We use the same argument as in the SU(2)-NATD case to determine the range of the $\rho$ coordinate, (see \cite{Lozano:2020sae} for more details). Namely, we impose that the quantity,
\begin{gather}
\label{boundedB2}
\frac{1}{4\pi^2}|\int_{\text{AdS}_2}	B_2|\in [0,1),
\end{gather}
is bounded and use a regularised volume for AdS$_2$\footnote{This reguralisation prescription is taken from \cite{Lozano:2020sae}.},
\begin{gather}
	\text{Vol}_{\text{AdS}_2}=4\pi^2.
\end{gather}

For $B_2$ in \eqref{NATD1} to satisfy \eqref{boundedB2} a large gauge transformation is needed as we move along $\rho$. Namely, for $\rho\in[\rho_k,\rho_{k+1}]$ we need to perform $B_2\to B_2+\pi k \text{vol}_{\text{AdS}_2}$, with
\begin{gather}
	\label{rhoLGT}
	\frac{\rho_k^3}{\rho_{k}^2-\rho_0^2}=2\pi k.
\end{gather}

We continue the study of the background \eqref{NATD1} by computing  the associated charges, obtained from the Page fluxes, defined by $\hat{F}=e^{-B_2}\wedge F$, given by, 
\begin{equation}
\begin{split}
\label{Page}
\hat{F}_0=&L^2,\quad\quad\quad
\hat{F}_2=-L^2k\pi\text{vol}_{\text{AdS}_2},\quad\quad\quad\quad
\hat{F}_4=-L^2 (M^4 \text{vol}_{\text{CY}_2}-2\rho\;\text{d}\rho\wedge\text{vol}_{\text{S}^3}),\\
\hat{F}_6=& L^2 (\pi kM^4 \text{vol}_{\text{CY}_2}+\rho(\rho-2\pi k)\;\text{d}\rho\wedge\text{vol}_{\text{S}^3})\wedge \text{vol}_{\text{AdS}_2},\\
\hat{F}_8=&2L^2M^4\rho\;\text{vol}_{\text{S}^3}\wedge \text{vol}_{\text{CY}_2} \wedge\text{d}\rho,\\
\hat{F}_{10}=&L^2M^4\rho(\rho-2\pi k)\text{vol}_{\text{AdS}_2}\wedge\text{vol}_{\text{S}^3}\wedge \text{vol}_{\text{CY}_2}\wedge\text{d}\rho,
\end{split}
\end{equation}
where we have taken into account the large gauge transformations $B_2\to B_2+\pi k \text{vol}_{\text{AdS}_2}$.
Inspecting the Page fluxes \eqref{Page}, we determine the type of branes that we have in the system. This is the  D0-D4-D$4'$-D8-F1 brane intersection depicted in Table \ref{brane-set-NATD}.
\begin{table}[h]
\begin{center}
\begin{tabular}{|c|c|c|c|c|c|c|c|c|c|c|}
\hline & $x^0$ & $x^1$ & $x^2$ & $x^3$ & $x^4$ & $x^5$ & $x^6$ & $x^7$ & $x^8$ & $x^9$\\ 
\hline
D0 & x & & & & & $$ & & &$$ &$$  \\
\hline
 D4 &  x &x &x  &x  &x  & & $$ & $$ & $$ &  \\
\hline
D$4'$ & x & $$ & $$ & $$ & $$ & $$ &x &x &x &x \\
\hline
D8 & x & x & x & x & x & & x & x & x &x  \\
\hline F1 &x & $$ & $$ & $$ & $$ & x& $$ & & & $$ \\ \hline
\end{tabular}
\caption{Brane set-up associated to our solution. Here x denotes the spacetime directions spanned by the various branes. $x^0$ corresponds to the time direction of the ten dimensional spacetime, $x^1, \dots , x^4$ are the coordinates spanned by the CY$_2$, $x^5$ is the direction where the F1-strings are stretched, and $ x^6, x^7, x^8, x^9$ are the coordinates where the SO$(4)$ symmetry is realised.}
\label{brane-set-NATD}
\end{center}
\end{table}

Using the expressions for the Page fluxes \eqref{Page} we compute the magnetic charges of Dp-branes using,
\begin{gather}
\label{magCha}
Q_{\text{Dp}}^m=\frac{1}{(2\pi)^{7-p}}\int_{\Sigma_{8-p}}\hat{F}_{8-p},
\end{gather}
where $\Sigma_{8-p}$ is a  $(8-p)$-dimensional manifold transverse to the directions of the Dp-brane. Furthermore, we define the electric charge of a Dp-brane as follows,
\begin{gather}
\label{eleCha}
Q_{\text{Dp}}^e=\frac{1}{(2\pi)^{p+1}}\int_{\text{AdS}_2\times\tilde{\Sigma}_{p}}\hat{F}_{p+2},
\end{gather}
here $\tilde{\Sigma}_p$ is defined as a $p-$dimensional manifold on which the brane extends. Both expressions, \eqref{magCha} and \eqref{eleCha} are written in units of $\alpha'=g_s=1$. 

As we anticipated, the background \eqref{NATD1} fits in the class of solutions presented in \cite{Lozano:2020sae}, that we briefly summarise in the next section. In such geometries, the D0 and D4-branes are interpreted in the dual field theory as instantons carrying electric charge. In turn, the D$4'$ and D8-branes have an interpretation as magnetically charged branes where the instantons lie. 
   In the interval $[\rho_k,\rho_{k+1}]$, these charges look in the following fashion,
\begin{equation}
\label{chargesNATD}
\begin{split}
	Q_{\text{D}8}^m&=2\pi\hat{F}_0=2\pi L^2,\\
	Q_{\text{D}4'}^m&=\frac{1}{(2\pi)^3}\int_{\text{CY}_2}\hat{F}_4=2\pi L^2M^4,\\
	 Q_{\text{D}0}^e&=\frac{1}{2\pi}\int_{\text{AdS}_2}\hat{F}_2=2\pi k L^2=k\;Q_{\text{D}8}^m,\\ 
	Q_{\text{D}4}^e&=\frac{1}{(2\pi)^5}\int_{\text{AdS}_2\times\text{CY}_2}\hat{F}_6=2\pi k L^2M^4=k\; Q_{\text{D}4'}^m\;,
\end{split}	
\end{equation}
%\textcolor{blue}{Additionally,
%\begin{equation}
%\begin{split}
%	&Q_{\text{D}0}^m=\frac{1}{(2\pi)^7}\int_{\Sigma_8}\hat{F}_8=\left(k+\frac{1}{2}\right)Q_{\text{D}4'}^m,\quad Q_{\text{D}4}^m=\frac{1}{(2\pi)^3}\int_{\text{S}^3\times \text{I}_\rho}\hat{F}_4=\left(k+\frac{1}{2}\right)Q_{\text{D}8}^m,\\
%	&Q_{\text{D}4'}^e=\frac{1}{(2\pi)^5}\int_{\text{AdS}_2\times\text{S}^3\times \text{I}_\rho}\hat{F}_6=\pi L^2\left(k+\frac{2}{3}\right),\quad  Q_{\text{D}8}^e=\frac{1}{(2\pi)^9}\int_{\text{AdS}_2\times\Sigma_8}\hat{F}_{10}=\pi L^2M^4\left(k+\frac{2}{3}\right),
%\end{split}	
%\end{equation}}
where we have used Vol$_{{\textrm{CY}}_2}=16\pi^4$. Furthermore, the fundamental strings are electrically charged with respect to the 3-form $H_3$,
\begin{gather}
Q_{\text{F1}}^e=\frac{1}{(2\pi)^2}\int_{\text{AdS}_2\times\text{I}_\rho}H_3=\left.\frac{1}{\pi}B_2\;\right|_{\rho_k}^{\rho_{k+1}}=1.
\end{gather}
One fundamental string is produced every time we cross the value $\rho=\rho_k$. Therefore in the interval $[0,\rho_k]$ there are $k$ F1-strings. 
%The charges in \eqref{chargesNATD} suggest that thought of a  NATD transformation the D1-D5 system is mapped to $k$ D0-D4 instantons at each interval that live in D$4'$-D8 branes.

\subsection{Holographic Central Charge}\label{HCCNATD}

In the spirit of the AdS/CFT correspondence, the study of AdS$_2$ geometries leads to consider one-dimensional dual field theories, where the definition of the central charge is subtle. In a conformal quantum mechanics the energy momentum tensor has only one component, and as the theory is conformal, it must vanish. We will interpret the central charge as counting the 
%follow the line studied in \cite{Lozano:2020txg, Lozano:2020sae, Lozano:2021rmk} in which the conformal quantum mechanics is considered to have many ground states and the central charge is interpreted as the 
number of vacuum states in the dual superconformal quantum mechanics, along the lines of \cite{Lozano:2020txg, Lozano:2020sae, Lozano:2021rmk}.    

%In \cite{Lozano:2020txg, Lozano:2020sae, Lozano:2021rmk} the central charge is interpreted as the number of vacuum states in the dual superconformal quantum mechanics and in the present work we will follow this line.  

We compute the holographic central charge following the prescription in \cite{Macpherson:2014eza, Bea:2015fja}, where this quantity is obtained from the volume of the internal manifold, accounting for a non-trivial dilaton,
\begin{gather}
\begin{split}
\label{hcc-NATD1}
&V_{int}=\int d^8x\;e^{-2\Phi}\sqrt{\det g_{8,ind}}=	2^5\pi^6L^4M^4\int_{\text{I}_\rho}(\rho^2-4L^4)\;\text{d}\rho,\\
&c_\text{hol}=\frac{3V_{int}}{4\pi G_N}=	\frac{3L^4M^4}{\pi}\int_{\text{I}_\rho}%_0^{2\pi(P+1)}
(\rho^2-4L^4)\;\text{d}\rho,
\end{split}
\end{gather}
where $G_N=8\pi^6$ in units $g_s=\alpha'=1$.
Since the dual manifold is non-compact the new background has an internal space of infinite volume that leads to an infinite holographic central charge, which points the solution needs a completion, as is shown in expression \eqref{hcc-NATD1}. %\textcolor{blue}{It also shows that a completion (as we show in Section \ref{field-theory}) can significantly define a holographic central charge.}  

 %However, the fact that the Lagrange multipliers take values in the Lie algebra suggest that global information of the dual background cannot be read into from the transformation itself. Thus,  we consider that internal geometry is globally unknown but compact. We will address this point  in Section \ref{field-theory} with a completion of the background \eqref{NATD1}.
%Since the dual manifold is uncompact the new background has an internal space of infinite volume, which leads to an infinite holographic central charge. However, the fact that the Lagrange multipliers take values in the Lie algebra suggest that the global information of the dual background cannot be read into from the transformation itself. Thus, we address this approach in Section \ref{field-theory} considering that internal geometry is globally unknown but compact.   

In the next section, we review the solutions constructed in \cite{Lozano:2020sae} in order to see that the background \eqref{NATD1} fits in that class of solutions. 
In turn, using the developments of \cite{Lozano:2020sae}, a concrete completion to the background \eqref{NATD1} generated by SL(2,$\mathbf{R}$)-NATD is proposed. Such completion in the geometry implies also a completion in the quiver, letting us to describe a well-defined CFT.

%In turn, using the developments of \cite{Lozano:2020sae}, we will propose a concrete completion to the background \eqref{NATD1}, generated by SL(2,$\mathbf{R}$)-NATD. Such completion in the geometry implies also a completion in the quiver, which allows describe a well-defined CFT.          
%%%%%%%%%%%%%%%%%%%%%%%%%%%%%%%%%%%%%%%%%%%%%%%%%%%%%
\section{The AdS$_2\times$S$^3\times$CY$_2$ solutions to massive IIA and their dual SCQM}\label{typeCreview}

In \cite{Lozano:2019emq} a  classification of AdS$_3\times$S$^2$ solutions to massive IIA supergravity with small (0,4) supersymmetry and SU(2)-structure was obtained.
%In \cite{Lozano:2019emq} a  classification of AdS$_3\times$S$^2$ solutions with small (0,4) supersymmetry to massive Type IIA supergravity with SU(2) structure and no electric NS flux was obtained. 
These solutions are warped products of the form AdS$_3\times$S$^2\times$M$_4\times$I preserving an SU(2) structure on the internal five-dimensional space. The M$_4$ is either a CY$_2$ or a 4d K\"{a}hler manifold. The respective classes of solutions are referred as class I and class II. 
In this section we briefly discuss the AdS$_2\times$S$^3\times$CY$_2$ solutions obtained via a double analytical continuation of the class I solutions above. These solutions were first constructed in \cite{Lozano:2020bxo} and then studied in detail in \cite{Lozano:2020sae}. These backgrounds are dual to SCQMs which were also studied in \cite{Lozano:2020sae}, and that we also review.
%\textcolor{blue}{As we will explain in Section \ref{QM-NATD} the background \eqref{NATD1} belongs to the geometries studied in \cite{Lozano:2020bxo,Lozano:2020sae}, which allows to propose a concrete completion for the solution \eqref{NATD1} and therefore a well-defined central charge.}
The study of the solutions constructed in \cite{Lozano:2020bxo,Lozano:2020sae} allows us to propose a concrete completion for the solution \eqref{NATD1} and therefore a well-defined central charge. We present the details of this completion in Section \ref{QM-NATD}.

%The study of the solutions constructed in \cite{Lozano:2020bxo,Lozano:2020sae} will allow us to propose a concrete completion for the solution \eqref{NATD1} and therefore a well-defined central charge.

A subset of the backgrounds studied in \cite{Lozano:2020bxo,Lozano:2020sae} --where we assume that the symmetries of the CY$_2$ are respected by the full solution--  read,
\begin{equation}\label{C-string}
	\begin{split}
	%\begin{center}
		&\text{d}s^2 = \frac{u}{\sqrt{h_4 h_8}} \left( \frac{h_4 h_8}{\Delta} \text{d}s^2_{\text{AdS}_2} + \text{d}s^2_{\text{S}^3}\right) + \sqrt{\frac{h_4}{h_8}} \text{d}s^2_{\text{CY}_2} + \frac{\sqrt{h_4 h_8}}{u} \text{d} \rho^2 \, ,\quad \Delta=4 h_4 h_8 - (u')^2, \\
		&e^{- 2\Phi}= \frac{h_8^{3/2}\Delta}{4 h_4^{1/2} u} \, ,\qquad\qquad 
		B_2 = - \frac{1}{2}\bigg( \rho -2\pi k+ \frac{u u'}{\Delta} \bigg)\text{vol}_{\text{AdS$_2$}},\, \\
		&\hat{F}_{0} = h_8' \, , \qquad\qquad \hat{F}_{2} =  -\frac{1}{2} \Big( h_8 -h_8'(\rho-2\pi k) \Big) \text{vol}_{\text{AdS$_2$}} \, , \\
		&\hat{F}_{4} = \left(2 h_8 \text{d} \rho  - \text{d} \bigg( \frac{u'u}{2 h_4} \bigg) \right) \wedge \text{vol}_{\text{S$^3$}}  - \partial_{\rho} h_4 \text{vol}_{\text{CY$_2$}} \,.
		%\end{center}
	\end{split}
\end{equation}
Here $\Phi$ is the dilaton and $B_2$ is the Kalb-Ramond field.  The warping functions $h_8$, $h_4$  and $u$ have support on $\rho$, with $u'=\partial_\rho u$. We have quoted the Page fluxes, $\hat{F}=e^{-B_2}\wedge F$, and included large gauge transformations\footnote{Like those that were studied in Section \ref{set-upNATD}.} of $B_2$ of parameter $k$, $B_2\to B_2+\pi k\text{vol}_{\text{AdS}_2}$.
The higher dimensional fluxes can be obtained as $F_{p} = (-1)^{[p/2]} \star_{10} F_{10-p}$.
Note that  $\Delta >0$, in order to guarantee a real dilaton and a metric with the correct signature. %The 2-form $H_2$ is defined in terms of the three functions $g_{1,2,3}$ and the vielbein on the CY$_2$,
%\begin{eqnarray*} \label{ges}
%H_2=g_1(\hat{e}^1\wedge \hat{e}^2-\hat{e}^3\wedge \hat{e}^4)+g_2(\hat{e}^1\wedge \hat{e}^3+\hat{e}^2\wedge \hat{e}^4)+g_3(\hat{e}^1\wedge \hat{e}^4-\hat{e}^2\wedge \hat{e}^3)	.
%\end{eqnarray*}

Supersymmetry holds whenever
%\begin{eqnarray}
%\label{susyC}
$u''=0$. 	
%\end{eqnarray}
%where $\hat{\star}_4$ is the Hodge dual on CY$_2$. 
 In turn, the Bianchi identities of the fluxes impose,
\begin{gather}
\begin{split}
h_8''=0\, , \qquad h_4''=0,%\mathrm{d}H_2=0, \qquad\frac{h_8}{u}\nabla_{\text{CY}_2}^2h_4+\partial_\rho^2h_4+\frac{2}{h_8^3}(g_1^2+g_2^2+g_3^2)=0,
\end{split}
\end{gather}
%away from localised sources.
%A further simplification consist in assuming the symmetries of the CY$_2$ are respected by the full solution, namely $H_2=0$ and $h_4=h_4(\rho)$. After this, the background reads,
%\begin{equation}\label{C-string}
%	\begin{split}
	%\begin{center}
%		&\text{d}s^2 = \frac{u}{\sqrt{h_4 h_8}} \left( \frac{h_4 h_8}{\Delta} \text{d}s^2_{\text{AdS}_2} + \text{d}s^2_{\text{S}^3}\right) + \sqrt{\frac{h_4}{h_8}} \text{d}s^2_{\text{CY}_2} + \frac{\sqrt{h_4 h_8}}{u} \text{d} \rho^2 \, ,\quad \Delta=4 h_4 h_8 - (u')^2, \\
%		&e^{- 2\Phi}= \frac{h_8^{3/2}\Delta}{4 h_4^{1/2} u} \, ,\qquad\qquad 
%		B_2 = - \frac{1}{2}\bigg( \rho -2\pi k+ \frac{u u'}{\Delta} \bigg)\text{vol}_{\text{AdS$_2$}},\, \\
%		&\hat{F}_{0} = h_8' \, , \qquad\qquad \hat{F}_{2} =  -\frac{1}{2} \Big( h_8 -h_8'(\rho-2\pi k) \Big) \text{vol}_{\text{AdS$_2$}} \, , \\
%		&\hat{F}_{4} = \left(2 h_8 \text{d} \rho  - \text{d} \bigg( \frac{u'u}{2 h_4} \bigg) \right) \wedge \text{vol}_{\text{S$^3$}}  - \partial_{\rho} h_4 \text{vol}_{\text{CY$_2$}} \,.
		%\end{center}
%	\end{split}
%\end{equation}
%Here we have quote the Page fluxes, $\hat{F}=e^{-B_2}\wedge F$, and include large gauge transformations of $B_2$ of parameter $k$, $B_2\to B_2+\pi k\;\text{vol}_{\text{AdS}_2}$. The background in \eqref{C-string} is a SUSY solution of the Massive IIA equations of motion if the warping functions satisfy,
%\begin{eqnarray}
%h_4''=0,\qquad h_8''=0,\qquad u''=0,
%\end{eqnarray}
 away from localised sources, which makes $h_8$ and $h_4$ are piecewise linear functions of $\rho$.

 %As it will be useful what follows we also obtain,
%\begin{eqnarray}
%H_7=-\star H_3=\frac{2u}{\sqrt{h_4h_8^5}}\left(h_4h_8-\frac{u u'\partial_\rho(h_4h_8)}{4 h_4 h_8 - (u')^2}\right)\text{vol}_{\text{S}^3}\wedge\text{vol}_{\text{CY}_2}
%\end{eqnarray}

Particular solutions were studied in \cite{Lozano:2020sae} where the functions $h_4$ and $h_8$ are piecewise continuous as follows,  
\begin{gather} \label{profileh4sp}
h_4(\rho)\!=%\!\Upsilon\! \,h_4(\rho)\!=\!\!
                    %\Upsilon\;\!\!
                    \left\{ \begin{array}{cccrcl}
                       \frac{\beta_0 }{2\pi}
                       \rho & 0\leq \rho\leq 2\pi, &\\
                                     \alpha_k\! +\! \frac{\beta_k}{2\pi}(\rho-2\pi k) &~~ 2\pi k\leq \rho \leq 2\pi(k+1),& ~~k=1,...,P-1\\
                      \alpha_P-  \frac{\alpha_P}{2\pi}(\rho-2\pi P) &~~ 2\pi P\leq \rho \leq 2\pi(P+1),&
                                             \end{array}
\right.\\
\label{profileh8sp}
h_8(\rho)
                    =\left\{ \begin{array}{cccrcl}
                       \frac{\nu_0 }{2\pi}
                       \rho & 0\leq \rho\leq 2\pi, &\\
                        \mu_k+ \frac{\nu_k}{2\pi}(\rho-2\pi k) &~~ 2\pi k\leq \rho \leq 2\pi(k+1),& ~~k=1,...,P-1\\
                      \mu_P-  \frac{\mu_P}{2\pi}(\rho-2\pi P) &~~ 2\pi P\leq \rho \leq 2\pi(P+1).&
                                             \end{array}
\right.
\end{gather}
For $u'=0$ the previous functions vanish at $\rho=0$ and $\rho=2\pi (P+1)$, where the space begins and ends. The $\alpha_k$, $\beta_k$, $\mu_k$ and $\nu_k$ are integration constants, which are determined by imposing continuity of the NS sector as,
\begin{equation}
\begin{split}
\mu_k=\sum_{j=0}^{k-1}\nu_j,\qquad\qquad &\alpha_k=\sum_{j=0}^{k-1}\beta_j.\\
\end{split}
\end{equation}

Using the piecewise functions \eqref{profileh4sp} and  \eqref{profileh8sp} in the $[\rho_k, \rho_{k+1}]$ interval and the definitions \eqref{magCha}-\eqref{eleCha}, the expressions for the charges are,
\begin{equation}
\begin{split}
Q_{\text{D0}}^e=h_8-(\rho-2\pi k)h_8'=\mu_k,\qquad\qquad &Q_{\text{D4}}^e=h_4-(\rho-2\pi k)h_4'=\alpha_k,\\
Q_{\text{D4}'}^m=2\pi h_4'=\beta_k,\qquad\qquad &Q_{\text{D8}}^m=2\pi h_8'=\nu_k	,\\
\end{split}
\end{equation}
and given that,
\begin{equation}
\begin{split}
\label{bianchiflavour}
\text{d}\hat{F}_0=h_8''\text{d}\rho, \qquad \qquad &\text{d}\hat{F}_4=h_4''\text{d}\rho\wedge\text{vol}_{\text{AdS}_2},
\end{split}
\end{equation}
with,
\begin{gather}
\label{h2p}
h_8''=\frac{1}{2\pi}\sum_{j=1}^P(\nu_{j-1}-\nu_j)\delta(\rho-2\pi j),\qquad h_4''=\frac{1}{2\pi}\sum_{j=1}^P(\beta_{j-1}-\beta_j)\delta(\rho-2\pi j),	
\end{gather}
there are D8 and D4$'$ brane sources localised in the $\rho$ direction. In turn, both d$\hat{F}_8$ and the vol$_{\text{S}^3}$ component of d$\hat{F}_4$ vanish identically, which implies that D0 and D4 branes play the  r\^ole of colour branes.
 The brane set-up associated to the solution \eqref{C-string} consists of a D0-F1-D4-D$4'$-D8 brane intersection, as depicted in Table \ref{brane-set-NATD}. 

In addition, in \cite{Lozano:2020sae} the number of vacua was computed.
For the solutions defined by the above functions, it was shown that the holographic central charge is given by,
\begin{equation}
\label{CC-TypeC}
c_{\text{hol,1d}}=\frac{3V_{int}}{4\pi G_N}=\frac{3}{4\pi}\frac{\text{Vol}_{\text{CY}_2}}{(2\pi)^4}\int_{0}^{2\pi(P+1)}(4h_4h_8-(u')^2)\;\text{d}\rho.	
\end{equation}

In the next section we briefly describe the SCQM proposed in \cite{Lozano:2020sae} in order to extract information about the field theory associated to the background \eqref{NATD1}.

%%%%%%%%%%%
\subsection{The dual superconformal quantum mechanics}\label{SCQM-typeC}

In \cite{Lozano:2020sae}, a proposal for the quantum mechanics living on the D0-D4-D$4'$-D8-F1 brane intersection was given in terms of an ADHM quantum mechanics that generalises the one discussed in \cite{Kim:2016qqs}. This quantum mechanics was interpreted as describing the interactions between instantons and Wilson lines in 5d gauge theories with 8 Poincar\'e supersymmetries living in D4-D8 intersections.  
The complete D0-D4-D$4'$-D8-F1 brane intersection was split into two subsystems, D4-D$4'$-F1 and D0-D8-F1, that were first studied independently.

Let us start considering the D4-D$4'$-F1 brane subsystem. This subsystem was interpreted as a BPS Wilson line in the 5d theory living on the D4-branes. When probing the D4-branes with fundamental strings, D$4'$-branes transverse to the D4-branes are originated. These orthogonal D$4'$-branes carry  a magnetic charge $Q_{\text{D}4'}^m=2\pi h_4'$ proportional to the number of fundamental strings dissolved in the world-volume of the D4$'$-branes. 
Additionally, the D4-branes can be seen as instantons in the world-volume of the D8-branes \cite{Douglas:1995bn}, where the D4-brane wrapped on the CY$_2$ can be absorbed by a D8-brane and converted into an instanton.

The D0-D8-F1 brane subsystem is distributed as the D4-D$4'$-F1 previous case. Here a Wilson line is introduced into the QM living on the D0-branes, in this case D8-branes are originated by probing D0-branes with fundamental strings. The number of fundamental strings dissolved in the worlvolume of D8-branes is in correspondence with the magnetic charge of the D8-branes,  $Q_{\text{D}8}^m=2\pi h_8'$. In terms of instantons, the D0-brane is absorbed by a D$4'$-brane and converted into an instanton.

\begin{figure}[t]
\centering
\includegraphics[scale=0.5]{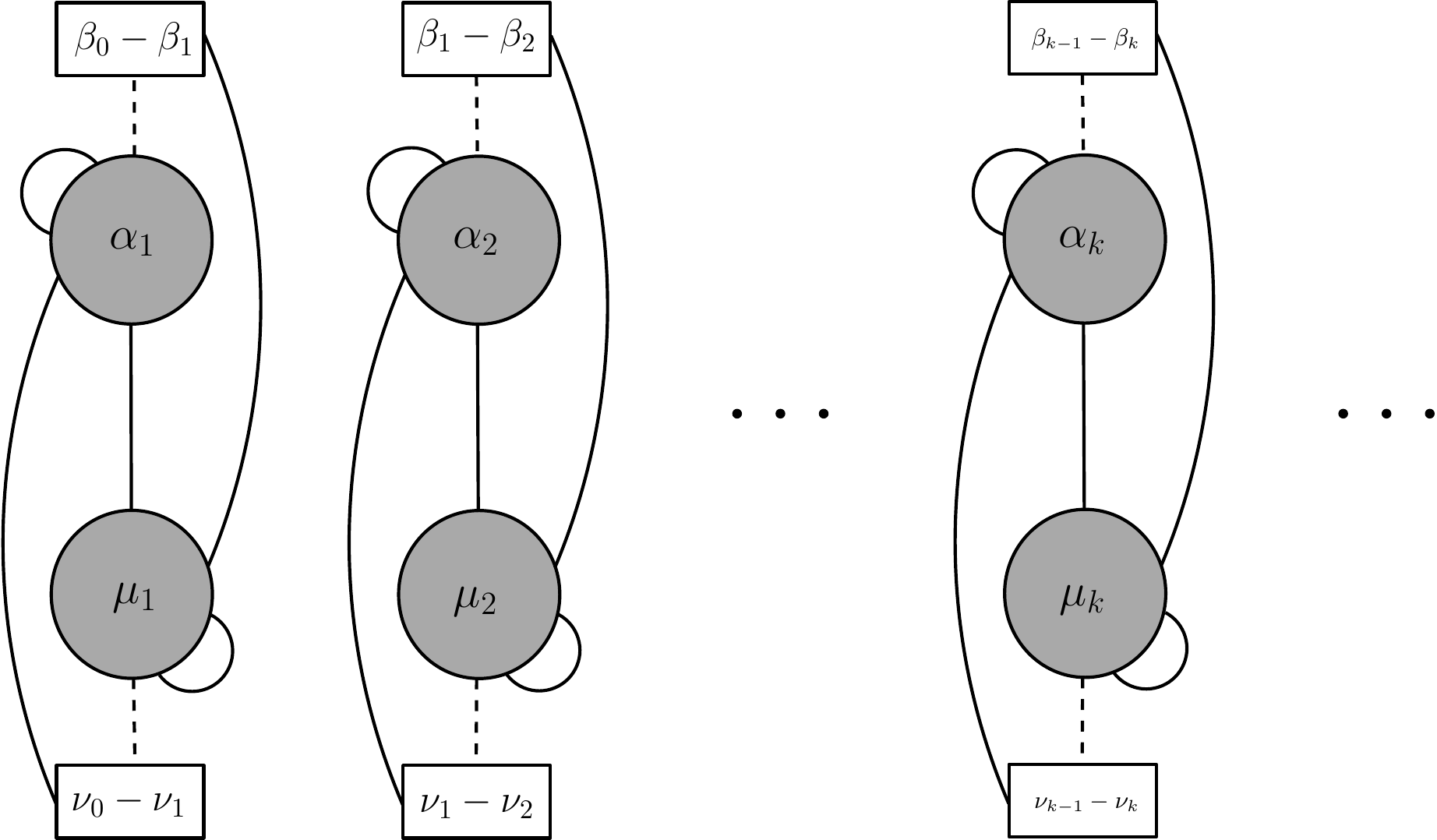}
\caption{A generic one dimensional quiver field theory whose IR limit is dual to the AdS$_2$ backgrounds given in \cite{Lozano:2020sae}.}
\label{QuiverGen}
\end{figure}
The proposal in \cite{Lozano:2020sae} is that the one dimensional $\mathcal{N}=4$ quantum mechanics living on the complete D0-D4-D$4'$-D8-F1 brane intersection describes the interactions between the two types of instantons and two types of Wilson loops in the $Q_{\text{D}4'}^m\times Q_{\text{D}8}^m$ antisymmetric representation of $\text{U}(Q_{\text{D}4}^e)\times \text{U}(Q_{\text{D}0}^e)$.

The SCQMs that live on these brane set-ups were analysed in \cite{Lozano:2020sae}. They are described in terms of a set of disconnected quivers as shown in Figure \ref{QuiverGen}, with gauge groups associated to the colour D0 and D4 branes (the latter wrapped on the CY$_2$) coupled to the D$4'$ and D8 flavour branes. The dynamics is described in terms of (4,4) vector multiplets, associated to gauge nodes (circles); (4,4) hypermultiplets in the adjoint representation connecting one gauge node to itself (semicircles in black lines); and (4,4) hypermultiplets in the bifundamental representation of the two gauge groups (vertical black lines). The connection with the flavour groups is through twisted (4,4)  bifundamental hypermultiplets, connecting the D0-branes with the D$4'$-branes and the D4-branes with the D8-branes (bent black lines), and (0,2) bifundamental Fermi multiplets, connecting the D4-branes with the D$4'$-branes and the D0-branes with the D8-branes (dashed lines) --see \cite{Lozano:2020sae} for more details.
\begin{figure}[t]
\centering
\includegraphics[scale=0.7]{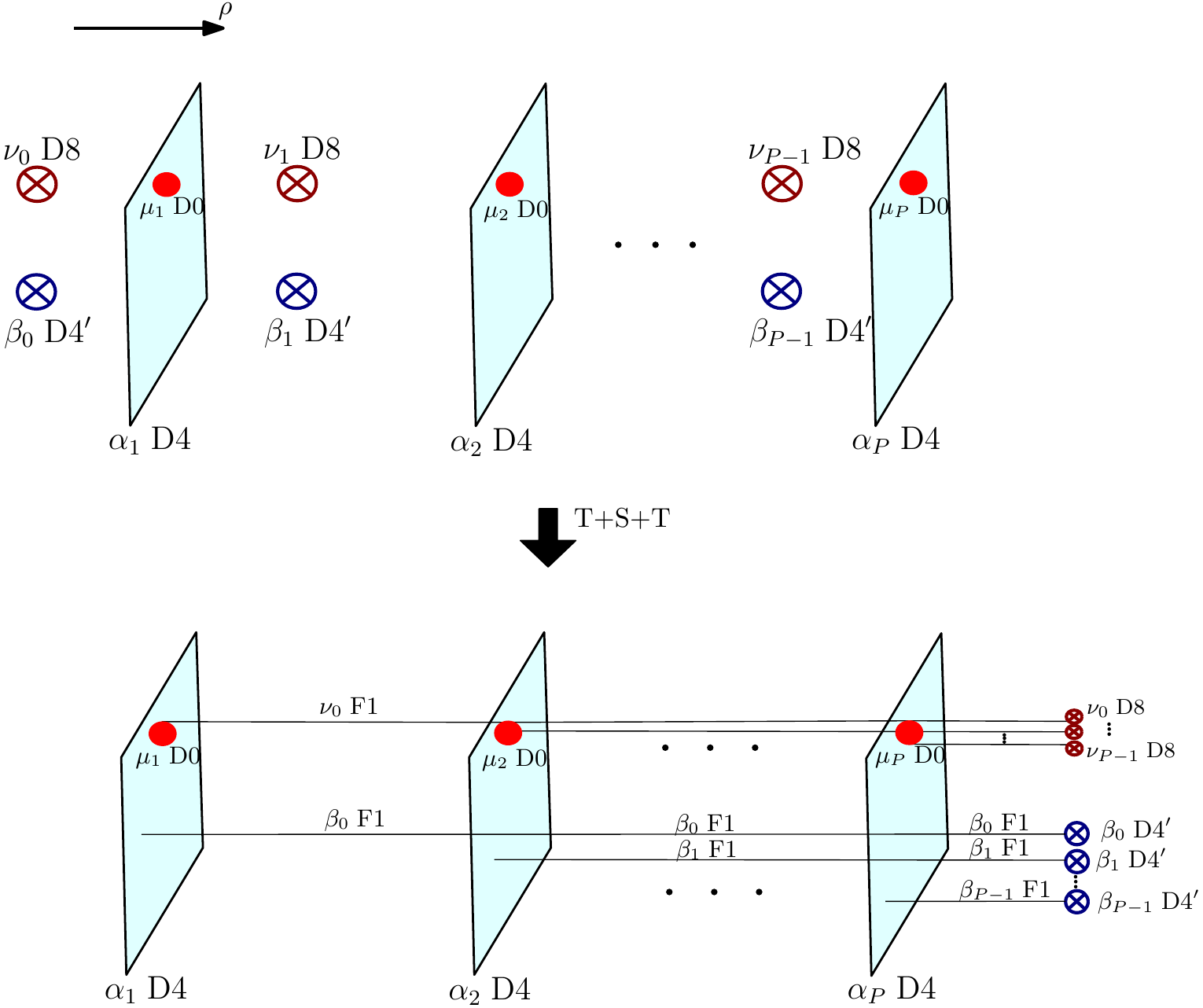}
\caption{(Top) Hanany-Witten like brane set-up associated with the quivers depicted in Figure \ref{QuiverGen}. Brane set-up equivalent to the previous one after a T+S+T duality transformation and Hanany-Witten moves (bottom).}
\label{HW-like} 
\end{figure}
 
 Such quivers, depicted in Figure \ref{QuiverGen}, can be read from the Hanany-Witten like brane set-up depicted at the top of Figure \ref{HW-like}. Here in each $[\rho_k, \rho_{k+1} ]$ interval there are $%Q_{\text{D}0}^e=
 \mu_k$ D0-branes and $%Q_{\text{D}4}^e=
 \alpha_k$ D4-branes, playing the r\^ole of colour branes. Orthogonal to them there are $%Q_{\text{D}8}^m=
 \nu_k$ D8-branes and $%Q_{\text{D}4'}^m=
 \beta_k$ D$4'$-branes, interpreted as flavour branes. In order to see the interpretation as Wilson lines one can proceed as follows (see \cite{Lozano:2020sae}). The D0-D4-D$4'$-D8-F1 brane set-up is taken to an F1-D3-NS5-NS7-D1 system in Type IIB through a T+S duality transformation. In this set-up, Hanany-Witten moves can be performed, which upon T-duality give the Type IIA configuration depicted at the bottom of Figure \ref{HW-like}. This configuration consists of $\nu_k$ coincident stacks of D8-branes and $\beta_k$ coincident stacks of D$4'$-branes, with $\mu_k$ and $\alpha_k$ F1-strings originating in the different $(\nu_0,\nu_1,...\nu_{k-1}; \beta_0,\beta_1,...,\beta_{k-1})$ coincident stacks of D8- and D$4'$-branes.
  The other endpoint of the F1-strings is on each stack of $\mu_k$ D0-branes  and $\alpha_k$ D4-branes. From this picture the description of Wilson loops in the $(\nu_0,\nu_1,...\nu_{k-1}; \beta_0,\beta_1,...,\beta_{k-1})$ completely antisymmetric representation of U$(\mu_k)$ and U$(\alpha_k)$, respectively, is recovered. In \cite{Lozano:2020sae}, this was interpreted as describing Wilson lines for each of the D0 and D4 gauge groups, given that they are in the completely antisymmetric representation they actually described backreacted D4-D0 baryon vertices \cite{Witten:1998xy} within the 5d CFT living in D4$'$-D8 brane intersections. The reader is referred to \cite{Lozano:2020sae} for more details on this construction.  
 
 In \cite{Lozano:2020sae}, it was shown that the holographic central charge (given by \eqref{CC-TypeC}), matches the field theory central charge, %from the quiver quantum mechanics that was described earlier, 
computed using the expression, 
 %In \cite{Lozano:2020sae}, the holographic central charge in \eqref{CC-TypeC} is matched with the field theory central charge computation from the quiver quantum mechanics that was described earlier. This expression looks in the following fashion,
 \begin{equation}
 \label{CC-ft}
 	c_{ft}=6(n_{hyp}-n_{vec}),
 \end{equation}
where $n_{hyp}$ counts the number of bifundamental, fundamental and adjoint (0,4) hypermultiplets and $n_{vec}$ counts the number of (0,4) vector multiplets, both in the UV description. 
The equation \eqref{CC-ft} was obtained in \cite{Putrov:2015jpa} for two-dimensional conformal field theories, and was  determined  by identifying the right-handed central charge with the U$(1)_R$ current two-point function.
With the expression \eqref{CC-ft}, both results, holographic and field theory central charge have been shown to agree for the 2d $\mathcal{N}=(0,4)$ quiver CFTs constructed in \cite{Lozano:2019jza, Lozano:2019zvg, Lozano:2019ywa}, as well as for the AdS$_2$/SCQM pairs proposed in \cite{Lozano:2020sae,Lozano:2020txg,Lozano:2021rmk}. 
In \cite{Lozano:2020txg},
 the agreement is kept since the one-dimensional quiver QMs are originated from the two-dimensional $\mathcal{N}=(0,4)$ CFTs upon dimensional reduction. However, in \cite{Lozano:2020sae,Lozano:2021rmk}, the equation \eqref{CC-ft} matches with the holographic result even though the 1d CFTs have not originated from the 2d ``mother'' CFTs. 
 
%Such an agreement is kept for the one-dimensional quiver quantum mechanics that originates from the two-dimensional $\mathcal{N}=(0,4)$ CFTs upon dimensional reduction \cite{Lozano:2020txg}. Further, the equation \eqref{CC-ft} matches with the 

%This was showed for 2d $\mathcal{N}=(0,4)$ quiver field theories constructed in \cite{Lozano:2019jza, Lozano:2019zvg, Lozano:2019ywa}. Further, for the AdS$_2$/SCQM pairs %of families 
%proposed in \cite{Lozano:2020sae,Lozano:2020txg,Lozano:2021rmk} this matching, between holographic and field theory central central, is valid.    

%Firstly, these results were studied in \cite{Lozano:2019jza, Lozano:2019zvg, Lozano:2019ywa} for the 2d $\mathcal{N}=(0,4)$ quiver field theories constructed there. In \cite{Lozano:2020txg}, it was shown that --through dimensional reduction-- both expressions are valid for their 1d $\mathcal{N}=4$ quiver quantum mechanics. Finally, the agreement is maintained for the quiver quantum mechanics that was described earlier and studied in detail in \cite{Lozano:2020sae}, as well for the 1d $\mathcal{N}=4$ quiver quantum mechanics discussed in \cite{Lozano:2021rmk}. 

As we anticipated, the background \eqref{NATD1} belongs to the classification provided in \cite{Lozano:2020sae}. Therefore, we will use the expression \eqref{CC-ft} to obtain the number of vacua of the superconformal quantum mechanics dual to our solution. The previous analysis guarantees its agreement with the holographic result.
 
After this summary, we turn to the solution \eqref{NATD1}, the main focus of this paper and show that it fits locally in the previous class of AdS$_2\times$S$^3\times$CY$_2\times$I solutions to Type IIA supergravity constructed in \cite{Lozano:2020sae}.   
% \textcolor{blue}{PONER LOS CAMPOS, SU DESCRIPCION, LOS QUIVERS, LA FIELD THEORY CENTRAL CHARGE}
 
 %\begin{itemize}
 %\item (4,4) vector multiplets, associated to gauge nodes (circles).
 %\item (4,4) hypermultiplets in the adjoint representation connecting one gauge node to itself (black lines). 
 %\item (4,4) hypermultiplets in the bifundamental representation	 of the two gauge groups (vertical black lines)
%\item twisted (4,4)  bifundamental hypermultiplets, coming  from the open strings that connect the D0-branes with the D$4'$-branes and the D4-branes with the D8-branes (bent black lines)
%\item (0,2) bifundamental Fermi multiplets, coming from the open strings that connect the D4- branes with the D$4'$-branes and D0-branes with the D8-branes (dash lines)
% \end{itemize}

%%%%%%%%%%%%%%%%%%%%%%%%%%%%%%%%%%%%%%%%%%%%%%%%%%%%%
\section{SCQM dual to the non-Abelian T-dual solution}\label{QM-NATD}

In this section we show that the solution \eqref{NATD1}, obtained as the SL($2,\mathbf{R}$)-NATD of the AdS$_3\times$S$^3\times$CY$_2$ solution to Type IIB supergravity, fits in the class of geometries  constructed in \cite{Lozano:2020sae}, that we have just reviewed. We will also provide a global completion to this solution by glueing it to itself.    

Consider the backgrounds \eqref{C-string}.
It is easy to see that the background \eqref{NATD1} fits locally in this class of %AdS$_2\times$S$^3\times$CY$_2$ 
solutions, with the simple choices, 
\begin{gather}
\begin{split}
\label{uNATD}
u= 4L^4 M^2 \rho,\qquad
h_4= L^2 M^4 \rho, \qquad
h_8=F_0\rho\, .
\end{split}
\end{gather}

In \cite{Lozano:2019ywa}, it was studied that the AdS$_3\times$S$^2\times$CY$_2\times$I solution constructed in \cite{Sfetsos:2010uq}, by acting SU(2)-NATD on the near horizon limit of the D1-D5 system, belongs to a subset of the geometries classified in \cite{Lozano:2019emq}. Therefore, since both classifications, \cite{Lozano:2019emq} and \cite{Lozano:2020bxo,Lozano:2020sae}, are related by a double analytical continuation, this fact strongly suggests that the background \eqref{NATD1} should be related to the solution obtained in \cite{Sfetsos:2010uq}, upon an analytical continuation prescription.
 
\begin{figure}[t]
	\centering
	\includegraphics[scale=0.8]{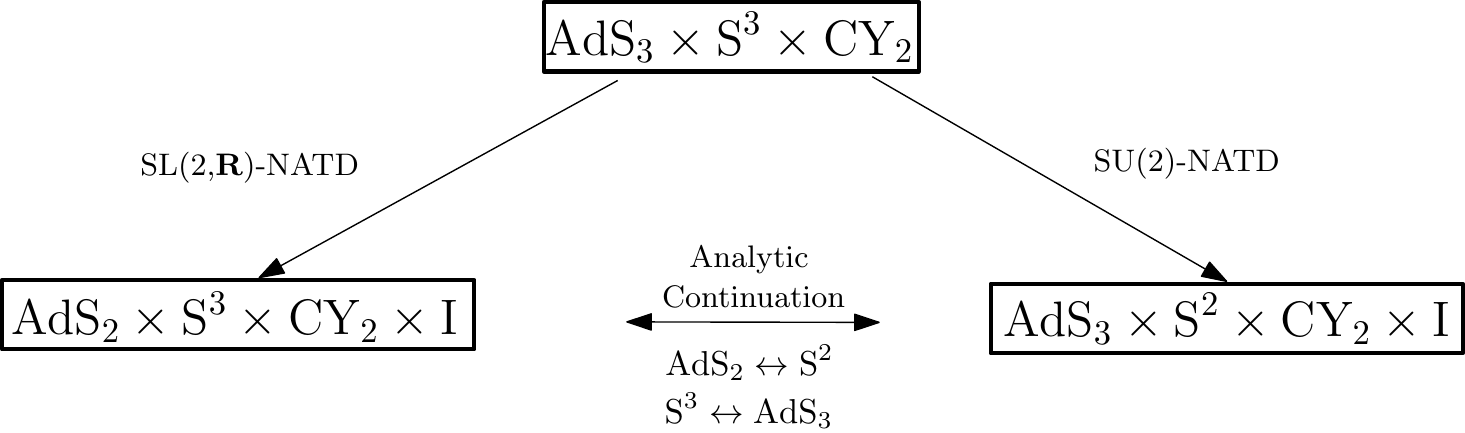}
	\caption{Relation between the solution \eqref{NATD1} and the solution obtained in \cite{Sfetsos:2010uq} through SU(2)-NATD.}
	\label{diagr}
\end{figure}

  This double analytical continuation works as follows, we focus on the Type IIA background given by \eqref{NATD1} and the AdS$_2$ and S$_3$ factors are interchanged as, 
  \begin{equation}
  	\begin{split}
  		&\text{d}s_{\text{AdS}_2}^2\to-\text{d}s_{\text{S}^2}^2,\qquad 
  		\text{d}s_{\text{S}^3}^2\to-\text{d}s_{\text{AdS}_3}^2. %\qquad  
  		%\rho\to i\rho,\qquad
  		%L\to i L,\qquad
  		%F_{i}\to -F_{i}.
  		\label{diagramaeq}
  	\end{split}
  \end{equation}
In order to get well-defined supergravity fields, we also need to analytically continue the following terms,   
%In addition, the relation with the particular AdS$_3\times$S$^2\times$CY$_2\times$I solution, in the classification of \cite{Lozano:2019emq} and constructed for the first time in \cite{Sfetsos:2010uq} by acting SU(2)-NATD on the AdS$_3\times$S$^3\times$CY$_2$ solution, is given by the following analytical continuation,
\begin{equation}
\begin{split}
%&\text{d}s_{\text{AdS}_2}^2\to-\text{d}s_{\text{S}^2}^2,\qquad 
%\text{d}s_{\text{S}^3}^2\to-\text{d}s_{\text{AdS}_3}^2, \qquad  
\rho\to i\rho,\qquad
L\to i L,\qquad
F_{i}\to -F_{i},
%\label{diagramaeq}
\end{split}
\end{equation}
where $F_i$ are the RR fluxes.
Thus, applying this set of transformations one finds the  AdS$_3\times$S$^2\times$ CY$_2\times$I solution to massive Type IIA supergravity with four Poincar\'e supersymmetries constructed for the first time in \cite{Sfetsos:2010uq}. We summarise these connexions in Figure \ref{diagr}.

\subsection{Completed NATD solution} \label{field-theory}

According to \eqref{profileh4sp}-\eqref{profileh8sp} one can choose a profile for the piecewise linear functions $h_4$, $h_8$ and propose a concrete way to complete the solution \eqref{NATD1}. In turn, completing the geometry implies a completion in the quiver, allowing us to match 
	between holographic and field theory computations.
%the field theory and holographic central charges.

We can complete the solution \eqref{NATD1} by terminating the $\rho$ interval at a certain value % that we take at
 $\rho_{2P}%=\rho_0\; (2P-1)
$ with $P\in \mathbb{Z}$\footnote{We choose the value $\rho_{2P}$ due to the completion is composed by two copies of the SL(2,$\mathbf{R}$)-NATD solution, glued between them. %Since we can define the SL(2,$\mathbf{R}$)-NATD solution up to $\rho_P$, the two copies go to $\rho_{2P}$. The subtraction  		
		%	in this completion we are glueing the SL(2,$\mathbf{R}$)-NATD solution to itself. That is, the SL(2,$\mathbf{R}$)-NATD solution provided by the functions \eqref{uNATD}
		%and we are subtracting the $\rho_0$ term in $\rho_{2P}$ because we are interested that the geometry --and therefore the piecewise linear functions-- vanish in $\rho_0$ in order to have the right signature.      
%In addition, we need the $\rho_P$ value to be an integer number.}
}.  Then, the piecewise functions \eqref{profileh4sp}-\eqref{profileh8sp} read,
\begin{gather}
	\label{profileuNATD}
	u=4L^4M^2 \rho  %~~~~~2\pi\tilde{\rho_0}\leq \rho \leq 2\pi\tilde{\rho_0} (2P-1)
	,
\end{gather}
\begin{gather} \label{profileh4NATD}
h_4(\rho)\!%=\!\Upsilon\! \,h_4(\rho)\!
=\!\!
                    L^2M^4\;\!\!\left\{ \begin{array}{cccrcl}
                      \rho
                        & \rho_0\leq \rho\leq \rho_P\;, &\\
                                     %\alpha_k\! +\! \frac{\beta_k}{2\pi}(\rho-2\pi k) &~~ 2\pi k\leq \rho \leq 2\pi(k+1),& ~~k=1,...,P-1\\
                     \rho_0- (\rho-\rho_{2P})&~~ \rho_P\leq \rho \leq \rho_{2P},&
                                             \end{array}
\right.\\
\label{profileh8NATD}
h_8(\rho)
=\!\!
                    L^2\;\!\!\left\{ \begin{array}{cccrcl}
                      \rho
                        & \rho_0\leq \rho\leq {\rho_P}, &\\
                                     %\alpha_k\! +\! \frac{\beta_k}{2\pi}(\rho-2\pi k) &~~ 2\pi k\leq \rho \leq 2\pi(k+1),& ~~k=1,...,P-1\\
                      {\rho_0}-  (\rho-\rho_{2P})&~~ {\rho_P}\leq \rho \leq\rho_{2P}.&
                                             \end{array}
\right.
\end{gather} 
%where $\tilde{\rho_0}$ is defined as $\tilde{\rho_0}=\frac{\rho_0}{2\pi}=\frac{L^2}{\pi}$. 
  The previous functions reproduce the behaviour \eqref{F1-singularity} for the metric and dilaton at both ends of the space and 
%With   we have a metric and dilaton well-defined, namely at beginning of the space, $\rho=2\pi\tilde{\rho_0}$, we recover the behaviour \eqref{F1-singularity}. At the end of the space, $\rho=2\pi\tilde{\rho_0}(2P-1)$, we obtain the same smeared F1-string singularity as \eqref{F1-singularity} but in this case $x=2\pi\tilde{\rho_0}(2P-1)-\rho$.  
one can check that the NS sector is continuous at $\rho_P$ when $\rho_P=\frac{\rho_0+\rho_{2P}}{2}%=\rho_0P
$. %when $n=P$.  
Hereinafter, we take the value $\rho_{2P}=\rho_0\; (2P-1)$ and use $\frac{\rho_0}{2\pi}$ dimensionless, namely $\frac{\rho_0}{2\pi}\to 1$, in order to obtain well-quantised charges. Thus, we get $\rho_P=2\pi P$.

Notice that the functions \eqref{profileh4NATD}-\eqref{profileh8NATD} are a simple example, with $\beta_k=\beta$ and $\nu_k=\nu$, for all intervals. This implies there are no flavour branes at the different intervals --with the exception of the [$\rho_{P-1},\rho_{P}$] interval, that we will analyse later. 

The Page fluxes \eqref{Page} %in both regions 
in each $[\rho_k,\rho_{k+1}]$ interval then read as follows%\footnote{Here we used $\frac{\rho_0}{2\pi}$ dimensionless, namely $\frac{\rho_0}{2\pi}\to 1$, in order to obtain well-quantised charges.
%associate the $n$ parameter in \eqref{profileh4NATD}-\eqref{profileh8NATD} with the $k$ parameter in the large gauge transformations. Strictly the relation between $n$ and $k$ should be, 
%\begin{equation*}
%	k=\tilde{\rho_0}n.
%\end{equation*}
%\label{footrho0}}
,
\begin{eqnarray} \label{profilepageF0}
&\hat{F}_0&\!%=\!\Upsilon\! \,h_4(\rho)\!
=\!\!
                    L^2\;\!\!\left\{ \begin{array}{cccrcl}
                      1
                        &%\rho_0\leq \rho\leq {\rho_k}& 
                        k=0,..., P-1, &\\
                              
                      -1&%~~ {\rho_k}\leq \rho \leq\rho_{2P-1}&
                      ~~~~k=P,..., (2P-1),&
                                             \end{array}
\right.\\
%\label{profilepageF0}
&\hat{F}_2&
=\!\!
                    \pi L^2\text{vol}_{\text{AdS}_2}\;\!\!\left\{ \begin{array}{cccrcl}
                      -k
                        &%\rho_0\leq \rho\leq {\rho_k}& 
                        k=0,..., P-1, &\\
                                     %\alpha_k\! +\! \frac{\beta_k}{2\pi}(\rho-2\pi k) &~~ 2\pi k\leq \rho \leq 2\pi(k+1),& ~~k=1,...,P-1\\
                      -(2P-k)&%~~ {\rho_k}\leq \rho \leq\rho_{2P-1}&
                      ~~~~k=P,..., (2P-1),&
                                             \end{array}
\right.\\
&\hat{F}_4^{\text{CY}_2}&
=\!\!
                    L^2M^4\text{vol}_{\text{CY}_2}\;\!\!\left\{ \begin{array}{cccrcl}
                      -1
                        &%\rho_0\leq \rho\leq {\rho_k}& 
                        ~~~~k=0,..., P-1, &\\
                                     %\alpha_k\! +\! \frac{\beta_k}{2\pi}(\rho-2\pi k) &~~ 2\pi k\leq \rho \leq 2\pi(k+1),& ~~k=1,...,P-1\\
                      1&%~~ {\rho_k}\leq \rho \leq\rho_{2P-1}&
                      ~~~~~~~~k=P,..., (2P-1),&
                                             \end{array}
\right.\\
&\hat{F}_6^{\text{CY}_2}&
=\!\!
                    \pi L^2M^4\text{vol}_{\text{AdS}_2}\wedge\text{vol}_{\text{CY}_2}\;\!\!\left\{ \begin{array}{cccrcl}
                      k
                        &%\rho_0\leq \rho\leq {\rho_k}& 
                        k=0,..., P-1, &\\
                                     %\alpha_k\! +\! \frac{\beta_k}{2\pi}(\rho-2\pi k) &~~ 2\pi k\leq \rho \leq 2\pi(k+1),& ~~k=1,...,P-1\\
                      (2P-k)&%~~ {\rho_k}\leq \rho \leq\rho_{2P-1}&
                      ~~~~k=P,..., (2P-1),&
                                             \end{array}
\right.
\end{eqnarray} 
Here we show the component over CY$_2$ for  $\hat{F}_4$ and $\hat{F}_6$.
The 2-form and 6-form Page fluxes are continuous at $\rho_P$ and the change of sign in the 0-form and 4-form Page fluxes is due to the presence of D8 and D$4'$ flavour branes at [$\rho_{P-1},\rho_{P}$] interval.

\begin{figure}[t]
	\centering
	\includegraphics[scale=0.7]{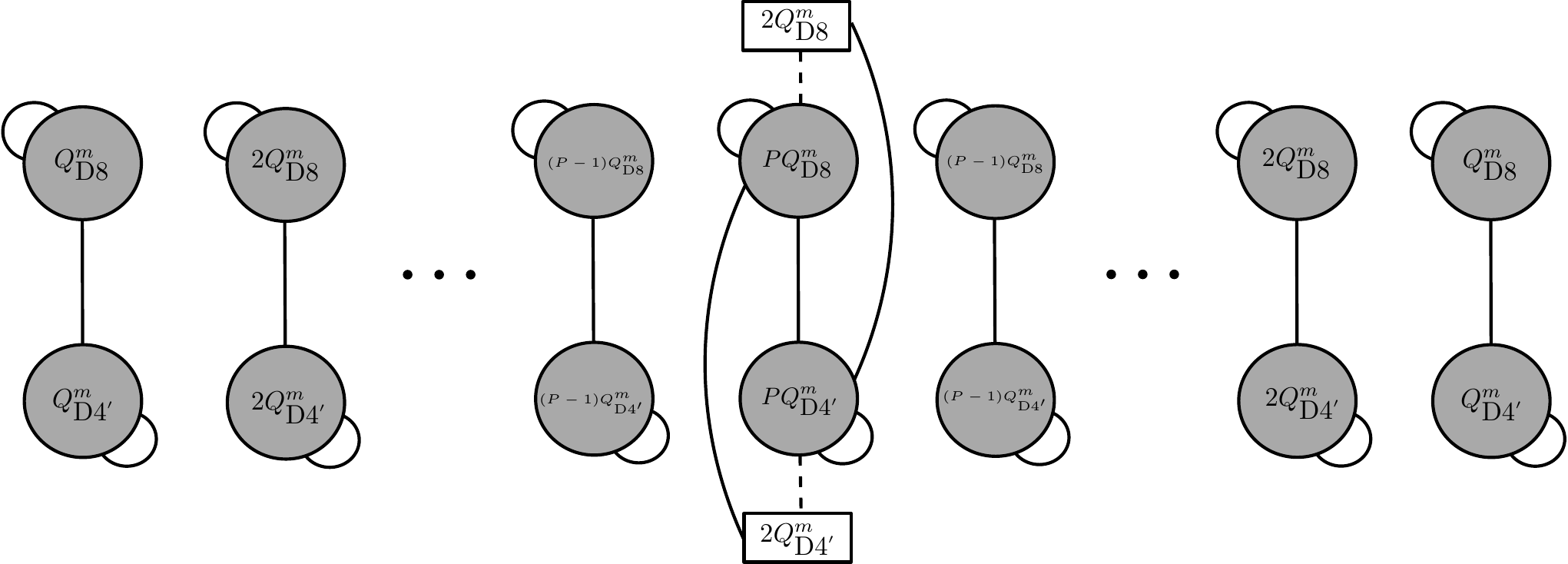}
	\caption{Symmetric completed quiver associated to the NATD solution.}
	\label{QuiverNATD}
\end{figure}
The corresponding quantised charges read,
\begin{eqnarray} \label{profilepageF0}
&Q_{\text{D8}}^m\!%=\!\Upsilon\! \,h_4(\rho)\!
&=\!\!
                    2\pi L^2\;\!\!\left\{ \begin{array}{cccrcl}
                      1
                        & %\rho_{k-1}\leq \rho\leq {\rho_k}&
                         k=0,..., P-1, &\\
                      -1&%~~ {\rho_{k-1}}\leq \rho \leq\rho_{k}&
                      ~~~~k=P,..., (2P-1),&
                                             \end{array}
\right.\\
%\label{profilepageF0}
&Q_{\text{D0}}^e
&=\!\!
                   Q_{\text{D8}}^m\;\!\!\left\{ \begin{array}{cccrcl}
                      -k
                        &%\rho_{k-1}\leq \rho\leq {\rho_k}& 
                        k=0,..., P-1, &\\
                                     %\alpha_k\! +\! \frac{\beta_k}{2\pi}(\rho-2\pi k) &~~ 2\pi k\leq \rho \leq 2\pi(k+1),& ~~k=1,...,P-1\\
                      -(2P-k)&%~~ {\rho_{k-1}}\leq \rho \leq\rho_{k}&
                      ~~~~k=P,..., (2P-1),&
                                             \end{array}
\right.\\
&Q_{\text{D4}'}^m
&=\!\!
                    2\pi L^2M^4\;\!\!\left\{ \begin{array}{cccrcl}
                      -1
                        &%\rho_{k-1}\leq \rho\leq {\rho_k}&
                         k=0,..., P-1, &\\
                                     %\alpha_k\! +\! \frac{\beta_k}{2\pi}(\rho-2\pi k) &~~ 2\pi k\leq \rho \leq 2\pi(k+1),& ~~k=1,...,P-1\\
                      1&%~~ {\rho_{k-1}}\leq \rho \leq\rho_{k}&
                      ~~~~k={P},..., (2P-1),&
                                             \end{array}
\right.\\
&Q_{\text{D4}}^e
&=\!\!
                   Q_{\text{D$4'$}}^m\;\!\!\left\{ \begin{array}{cccrcl}
                      k
                        &%\rho_{k-1}\leq \rho\leq {\rho_k}& 
                        k=0,..., P-1, &\\
                                     %\alpha_k\! +\! \frac{\beta_k}{2\pi}(\rho-2\pi k) &~~ 2\pi k\leq \rho \leq 2\pi(k+1),& ~~k=1,...,P-1\\
                      (2P-k)&%~~ {\rho_{k-1}}\leq \rho \leq\rho_{k}&
                      ~~~~k=P,..., (2P-1).&
                               \end{array}
\right.
\end{eqnarray}
%\begin{eqnarray*} 
%\begin{tabular}{|c||c|}
%\hline 
%$2L^2\leq \rho\leq L^2(P+1)$ & $L^2(P+1)\leq \rho \leq 2L^2P$ \\
% $1\leq k\leq\frac{P+1}{2}$ &$\frac{P+1}{2}\leq k\leq P$ 
%\\ 
%\hline\hline
% $Q_{\text{D8}}^m=2\pi L^2$ & $Q_{\text{D8}}^m=-2\pi L^2$  \\
%\hline
% $Q_{\text{D0}}^e=-k\;Q_{\text{D8}}^m$ &$Q_{\text{D0}}^e=(P+1-k)Q_{\text{D8}}^m$   \\
%\hline
%$Q_{\text{D$4'$}}^m=-2\pi L^2M^4$& $Q_{\text{D$4'$}}^m=2\pi L^2M^4$   \\
%\hline 
%$Q_{\text{D4}}^e=-k\;Q_{\text{D$4'$}}^m$ & $Q_{\text{D4}}^e=(P+1-k)Q_{\text{D$4'$}}^m$   \\ \hline
%\end{tabular}
%\end{center}
%\end{eqnarray*}
Thus, the D0 and D4 brane charges increase linearly in the $0\leq k\leq P$ region, and decrease linearly in the $P+1\leq k\leq 2P-1$ region, until the value $k=2P-1$ is reached. %,  where the geometry terminates. 
 Here the minus sign in the charges denotes anti-Dp brane charge. The quiver for the configuration \eqref{profileh4NATD}-\eqref{profileh8NATD} is depicted in Figure \ref{QuiverNATD}.

The discontinuities at $\rho_P$ are translated into $2Q_{\text{D$4'$}}^m$ and $2Q_{\text{D8}}^m$ flavour groups according to,
\begin{eqnarray}
N_{\text{D4}'}^{\left[P-1,P\right]}&=&\frac{1}{(2\pi)^3}\int_{\text{CY}_2}\hat{F}_4=\frac{1}{(2\pi)^3}\int_{\text{CY}_2\times \text{I}_\rho}\text{d}\hat{F}_4
%=\frac{1}{(2\pi)^3}\int_{\text{T}^4\times \text{I}_\rho}h_4''\text{d}\rho\wedge\text{vol}_{\text{T}^4}
=\beta_{P-1}-\beta_{P}
%=2\pi J^2M^4-(-2\pi L^2M^4)
=2Q_{\text{D$4'$}}^m\\
N_{\text{D8}}^{\left[P-1,P\right]}&=&2\pi\hat{F}_0=2\pi\int_{\text{I}_\rho}\text{d}\hat{F}_0
%=\frac{1}{(2\pi)^3}\int_{\text{T}^4\times \text{I}_\rho}h_4''\text{d}\rho\wedge\text{vol}_{\text{T}^4}
=\nu_{P-1}-\nu_{P}
%=2\pi J^2M^4-(-2\pi L^2M^4)
=2Q_{\text{D8}}^m
\end{eqnarray}
where we used the expressions \eqref{bianchiflavour}-\eqref{h2p} with $\beta_{P-1}=2\pi L^2M^4$, $\beta_{P}=-2\pi L^2M^4$, $\nu_{P-1}=2\pi L^2$ and $\nu_{P}=-2\pi L^2$.

\begin{figure}[t]
	\centering
	\includegraphics[width=16cm, height=6cm]{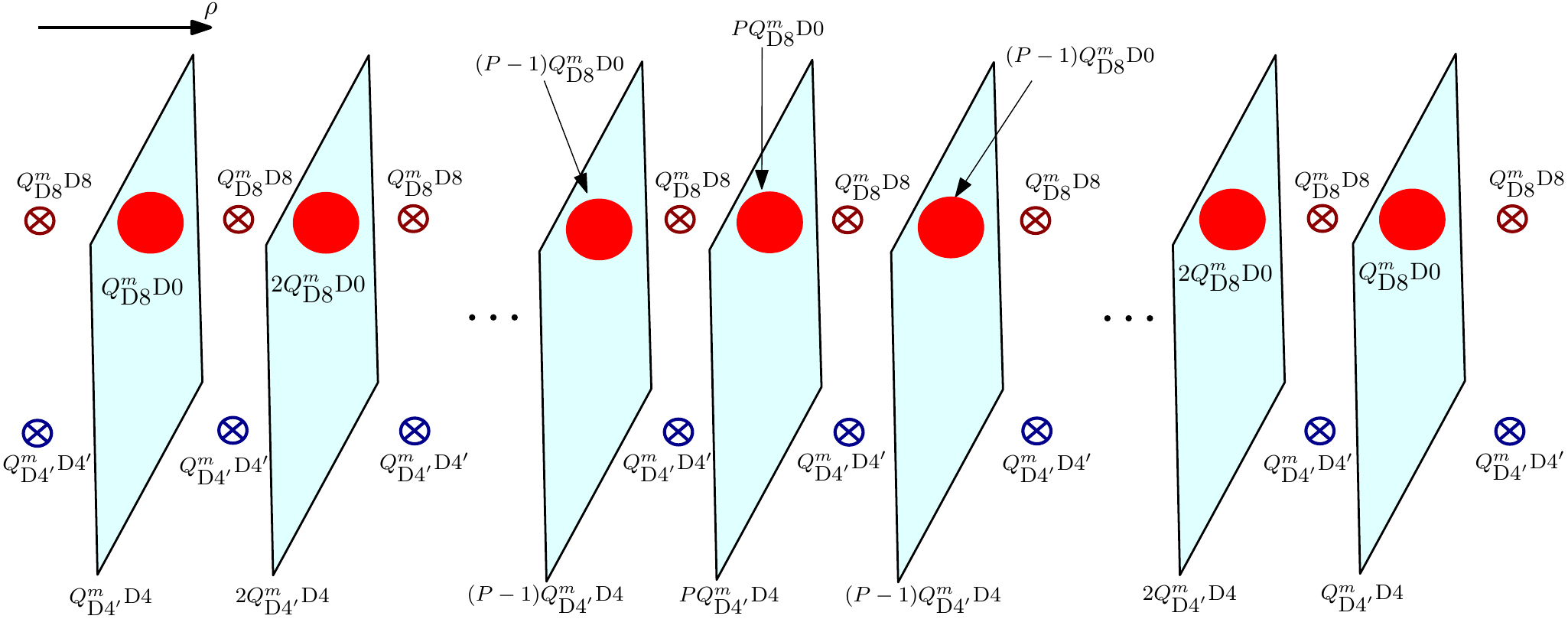}
	\caption{The Hanany-Witten like brane set-up for the completed non-Abelian T-dual solution, underlying the quiver depicted in Figure \ref{QuiverNATD}. 
	}
	\label{HW-like-NATD} 
\end{figure}

The quiver shown in Figure \ref{QuiverNATD} can be translated to the description reviewed in Section \ref{SCQM-typeC}. The Hanany-Witten like brane set-up is shown in Figure \ref{HW-like-NATD}. In each $[\rho_{k},\rho_{k+1}]$ interval, for $k=0,...,P-1$, we have $k Q_{\textrm{D8}}^m$ D0-branes and  $k Q_{\textrm{D4}'}^m$ D4-branes. For $k=P,...,2P-1$ we have  $(2P-k) Q_{\textrm{D8}}^m$ D0-branes and $(2P-k) Q_{\textrm{D4}'}^m$ D4-branes. Orthogonal to them, in each interval, there are $Q_{\textrm{D8}}^m$ D8-branes and $Q_{\textrm{D4}'}^m$ D4$'$-branes, playing the r\^ole of flavour branes.

%As proposed in \cite{Lozano:2020sae} and we reviewed in Section \ref{SCQM-typeC}, one can lead the D0-D4-D4$'$-D8-F1 system to F1-D3-NS5-NS7-D7 brane setup in Type IIB  through  a T-S duality transformation, where the Hanany-Witten moves are carry out. When we back to Type IIA the configuration in Figure \ref{HW-like-NATDII} is obtained.

\begin{figure}[t]
	\centering
	\includegraphics[width=16cm, height=8cm]{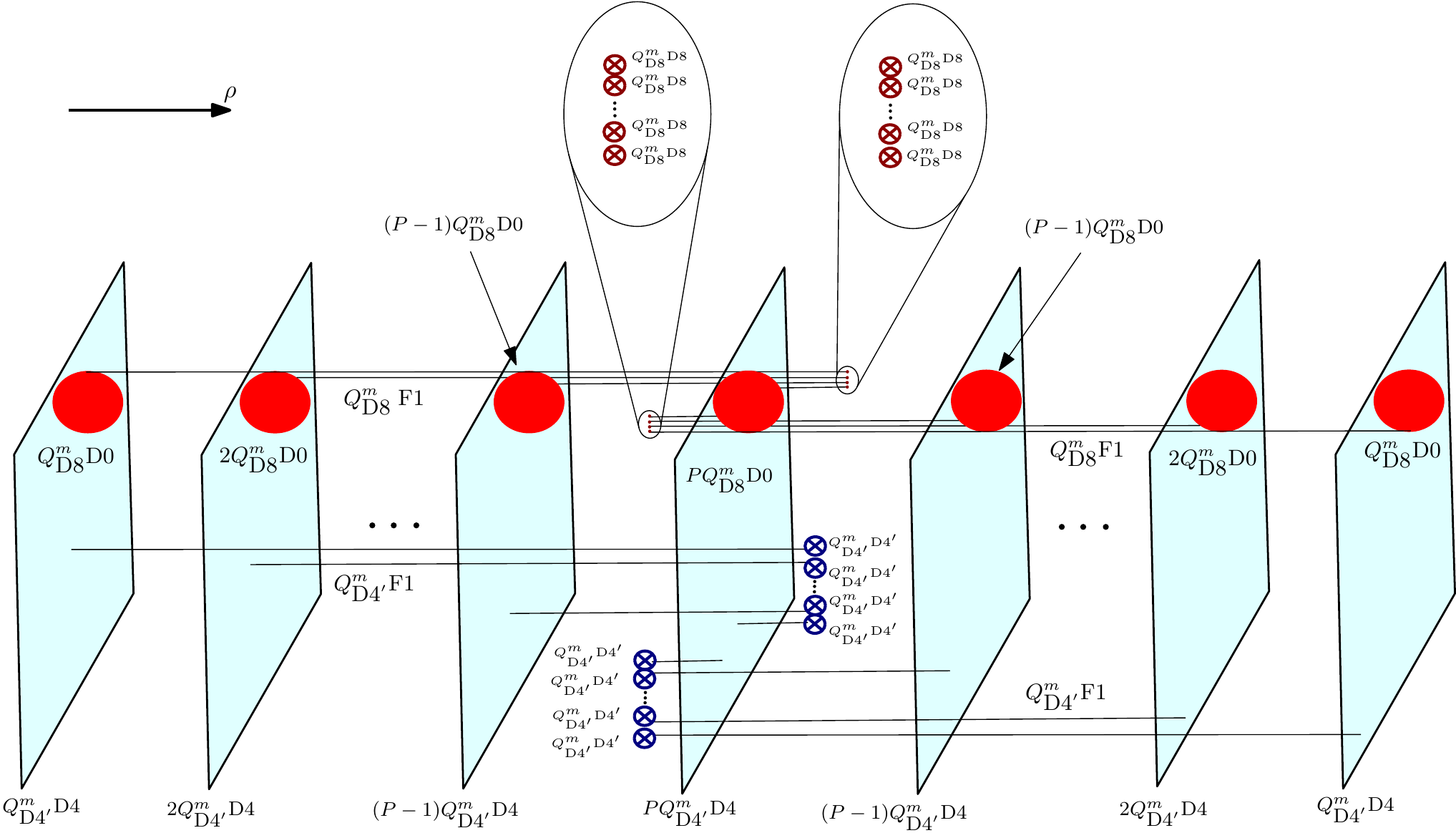}
	\caption{Symmetric brane set-up after a T+S+T duality transformation and Hanany-Witten moves from the brane set-up depicted in Figure \ref{HW-like-NATD}.
	}
	\label{HW-like-NATDII} 
\end{figure}

As proposed in \cite{Lozano:2020sae} and we reviewed in Section \ref{SCQM-typeC}, one can perform a T-S-T duality transformation\footnote{That is a T-(S-duality)-T transformation.} to the D0-D4-D4$'$-D8-F1 system.
Consider the left-hand side of the Hanany-Witten like brane set-up shown in Figure \ref{HW-like-NATD}, from the first $Q_{\textrm{D8}}^m\text{D}8$- and $Q_{\textrm{D4}'}^m\text{D}4'$-branes  until the $PQ_{\textrm{D8}}^m\text{D}0$- and $PQ_{\textrm{D4}'}^m\text{D}4$-branes. It is easy to see that this subsystem is equivalent to the brane set-up depicted on the top of Figure \ref{HW-like} (with $\nu_i=Q_{\textrm{D8}}^m$, $\beta_i=Q_{\textrm{D4}'}^m$, $\mu_j=jQ_{\textrm{D8}}^m$ and $\alpha_j=jQ_{\textrm{D4}'}^m$, for $i=0,1,...,P-1$ and $j=1,...,P$). When we perform the T+S+T transformation on the left-hand configuration an equivalent system to the bottom in Figure \ref{HW-like} is obtained. This is depicted on the left-hand side in Figure \ref{HW-like-NATDII}, here the now coincident D8-branes and D4$'$-branes are to the right of the $PQ_{\textrm{D8}}^m$D0 and $PQ_{\textrm{D4}'}^m$D4 stacks. 
On the right-hand side of the Hanany-Witten like brane set-ups, shown in Figure \ref{HW-like-NATD} and Figure \ref{HW-like-NATDII}, we have the same configuration that on the left-hand side, since the right-hand side is the symmetric part of the left-hand side. That is, the complete configuration is the left-hand side glued to itself.
 %the left configuration in Figure \ref{HW-like-NATD} after the T+S+T transformation is equivalent to the system on the bottom of Figure \ref{HW-like} 

%Thus, after the T+S+T transformation the configuration on the left hand side in Figure \ref{HW-like-NATDII} is obtained the D8 and D4$'$ are led to the right of the $PQ_{\textrm{D8}}^m$D0- and $PQ_{\textrm{D4}'}^m$D4-branes, where we now get coincident D8-branes and coincident D4$'$-branes.

%The quiver shown in the Figure \ref{QuiverNATD} can be translate to the description reviewed in section \ref{SCQM-typeC}. The Hanany-Witten like brane set-up is shown on the top in the Figure \ref{HW-like-NATD}.

%\textcolor{blue}{Let us focus on the left hand side of the Hanany-Witten brane set-up shown in Figure \ref{HW-like-NATDII}, from $Q_{\textrm{D8}}^m\text{D}0$ and $Q_{\textrm{D8}}^m\text{D}0$  until $PQ_{\textrm{D8}}^m\text{D}0$ and $PQ_{\textrm{D8}}^m\text{D}0$.}  

%Here we have carried --with the Hanany-Witten moves--  the D8 and D4$'$ branes around the stacks with $PQ_{\textrm{D8}}^m\text{D}0$ and $PQ_{\textrm{D4}'}^m\text{D}4$ branes, showing a symmetric brane set-up.

Let us focus on the D0-D8-F1 system on the left-hand side of the Hanany-Witten like brane set-up shown in Figure \ref{HW-like-NATDII} (from $Q_{\textrm{D8}}^m\text{D}0$  until $PQ_{\textrm{D8}}^m\text{D}0$)\footnote{Since on the right-hand side we have the same configuration.}. After the T+S+T transformation, we obtain $P$ stacks of $Q_{\textrm{D8}}^m$D8-branes --depicted in Figure \ref{HW-like-NATDII} to the right of the $PQ_{\textrm{D8}}^m$D0-branes--  
with $PQ_{\textrm{D8}}^m$F1-strings originating in the different coincident stacks of D8-branes. The other endpoint of the F1-strings is on each stack of $kQ_{\textrm{D8}}^m$D0-branes. For the D4-D4$'$-F1 system we have a similar configuration, namely  $P$ stacks of $Q_{\textrm{D4}'}^m$D4$'$-branes, with $PQ_{\textrm{D4}'}^m$F1-strings attached to them. These F1-strings have the other end point on the different $kQ_{\textrm{D4}'}^m$ stacks of D4-branes. Thus,
as we reviewed in Section \ref{SCQM-typeC}, the system can be interpreted as %D0 and D4 instantons in the world-volumes of D8 and D4$'$ branes interacting with 
Wilson loops %\footnote{As we pointed out we can interpret this behaviour as a baryon vertex, in this case for the gauge groups U$(kQ_{\textrm{D8}}^m)\times$U$(kQ_{\textrm{D4}'}^m)$.} baryon vertices 
 in the $Q_{\textrm{D8}}^m\times Q_{\textrm{D4}'}^m$ completely  antisymmetric representation of the gauge groups U$(kQ_{\textrm{D8}}^m)\times$U$(kQ_{\textrm{D4}'}^m)$, that we interpret as describing the backreaction of the D4-D0 baryon vertices of a D4$'$-D8 brane intersection. 
 
  To be concrete, consider the SCQM that arises in the very low energy limit of a D4$'$-D8 brane intersection, dual to a 5d QFT, where D4- and D0-brane baryon vertices are introduced. Namely, D4-brane (D0-brane) baryon vertices are linked to D4$'$-branes (D8-branes) with fundamental strings. In the IR these branes change their r\^ole, that is the gauge symmetry on both D4$'$- and D8-branes becomes global, shifting D4$'$ and D8 from colour to flavour branes and the D0- and D4-branes play now the r\^ole of colour branes of the backreacted geometry.

Furthermore, with the piecewise linear functions \eqref{profileuNATD}, \eqref{profileh4NATD} and \eqref{profileh8NATD} we can compute the holographic central charge\footnote{We used $\frac{\rho_0}{2\pi}\to 1$ as explained above.},
\begin{gather}
	\begin{split}
\label{chol}
	c_{hol}&=\frac{3}{\pi}L^4M^4\left(\int_{\rho_0}^{\rho_P}(\rho^2-\rho_0^2)\;\text{d}\rho+\int_{\rho_P}^{\rho_{2P}}\left(\left(\rho_0-(\rho-\rho_{2P})\right)^2-\rho_0^2\right)\;\text{d}\rho\right)\\
	&= Q_{\text{D4}'}^m Q_{\text{D8}}^m(4P^3-12P+8).
\end{split}
\end{gather}

In order to compare this result with the field theory computation in \eqref{CC-ft}, we need to compute the number of hypermultiplets and vector multiplets. For the quiver in Figure \ref{QuiverNATD} we obtain,
\begin{gather}
	\begin{split}
	n_{hyp}&=\left((Q_{\text{D$4'$}}^m)^2+(Q_{\text{D8}}^m)^2+Q_{\text{D4}'}^mQ_{\text{D8}}^m\right)\left(P^2+2\sum_{i=1}^{P-1}i^2\right),	\\
	n_{vec}&=\left((Q_{\text{D$4'$}}^m)^2+(Q_{\text{D8}}^m)^2\right)\left(P^2+2\sum_{i=1}^{P-1}i^2\right).
		\end{split}
\end{gather}
Thus, when the sums are performed we get the following expression for the field theory central charge,
\begin{eqnarray}
\begin{split}	
\label{cft}
	c_{ft}=6(n_{hyp}-n_{vec})&=6 Q_{\text{D4}'}^mQ_{\text{D8}}^m\left(P^2+2\sum_{i=1}^{P-1}i^2\right)\\
	&=  Q_{\text{D4}'}^mQ_{\text{D8}}^m (4P^3+2P).
	\end{split}
\end{eqnarray}
 We see that at large $Q_{\text{D8}}^m$, $Q_{\text{D4}'}^m$ and $P$ (in the holographic limit, which is long quivers with large ranks) the results \eqref{chol} and \eqref{cft} coincide.

\section{Conclusions}\label{conclusions}
In this paper we developed the implementation of NATD in supergravity backgrounds supporting an SL(2,$\mathbf{R}$) subgroup as part of their full isometry  group. Namely, we implemented the solution generating technique in non-compact spaces exhibiting an SO(2,2) $\cong$ SL(2,$\mathbf{R})_L\times$SL(2,$\mathbf{R})_R$ isometry group geometrically realised by an  AdS$_3$ space.
After the dualisation, the resultant dual geometry exhibits an SL(2,$\mathbf{R}$) isometry reflected geometrically as an AdS$_2$ space plus a non-compact new direction in the internal space. %\textcolor{blue}{The last is a consequence of the dualisation procedure, the original coordinates are replaced in the dual by coordinates living in the Lie algebra.}%-- even if the group used to construct the NATD is compact.
This non-compact direction arises since the Lagrange multipliers live in the Lie algebra of the SL(2,$\mathbf{R}$) group, which is by construction a vector space, $\mathbf{R}^3$. That is, the space dual to AdS$_3$ is locally AdS$_2\times\mathbf{R}^+$. %therefore the global information of the background cannot be inferred from the transformation itself \textcolor{blue}{The fact that the Lagrange multipliers take values in the Lie algebra suggest that global information of the dual background cannot be read into from the transformation itself. Thus,  we consider that internal geometry is globally unknown but compact. In this vein, a completion is needed for our dual background. We will tackle this point in Section \ref{field-theory} but advance that at both ends of the space the configuration is identified as a smeared fundamental string given by the expression \eqref{F1-singularity}.}      Due to the Lagrange multipliers live in the Lie algebra of the SL(2,$\mathbf{R}$) group, the global information of the background cannot be inferred from the transformation itself. In this way, a completion for the solution given in \eqref{NATD1} is needed. In Section \ref{field-theory}, we will provide a concrete completion, where at both ends of the space a smeared fundamental string,  like \eqref{F1-singularity}, is identified.

%After the recognisation, the resultant  geometry exhibits an spectator SL(2,$\mathbf{R}$) isometry reflected at level of the metric as an AdS$_2$ space.  In turn, the original coordinates are replaced in the dual by coordinates living in the Lie algebra therefore the dualisation procedure (even if the group used to construct the NATD is compact) generates an non-compact internal space. 

%  the dualisation procedure generates an non-compact internal space, which is associated that some of the new variables in the dual live in the Lie algebra, 
%. 
%as a by-product where even if the group used to construct the NATD were compact, some of the new variables, which now live in the Lie algebra, are not. 

We worked out in detail the SL(2,$\mathbf{R}$)-NATD solution of the AdS$_3\times$S$^3\times$CY$_2$ solution that arises in the near horizon limit of the D1-D5 brane intersection.
We found that the SL(2,$\mathbf{R}$)-NATD solution is a simple example in the classification constructed in \cite{Lozano:2020sae}. Further, our %AdS$_2\times$S$^3\times$CY$_2\times$I
background \eqref{NATD1} is related through an analytic continuation prescription to the AdS$_3\times$S$^2\times$CY$_2\times$I solution obtained in \cite{Sfetsos:2010uq}, as one of the first examples of AdS backgrounds generated through SU(2) non-Abelian T-duality.

An important drawback of non-Abelian T-duality is the lack of global information about the dual geometry, which cannot be inferred from the transformation itself. For this we used the fact that our solution \eqref{NATD1} fits in the classification constructed in \cite{Lozano:2020sae} which allowed us to  propose an explicit completion for the geometry. Unlike the two completions worked out in \cite{Lozano:2019ywa} for the SU(2)-NATD solution constructed therein, continuity of the NS sector allows only one possible completion for the geometry given in \eqref{NATD1}.
Our completion, shown in Section \ref{field-theory}, is obtained by glueing the SL(2,$\mathbf{R}$)-NATD solution to itself.  
We proposed a well-defined quiver quantum mechanics, dual to our AdS$_2$ solution, that flows in the IR to a superconformal quantum mechanics (based on the Hanany-Witten brane set-ups and Page charges), which admits an interpretation in terms of %interactions between instantons and Wilson loops in the $Q_{\textrm{D8}}^m\times Q_{\textrm{D4}'}^m$ antisymmetric representation of the gauge groups U$(kQ_{\textrm{D8}}^m)\times$U$(kQ_{\textrm{D4}'}^m)$.
 backreacted D4-D0 baryon vertices within the 5d QFT living in a D4$'$-D8 brane intersection.  In support of our proposal we checked the agreement between the holographic and field theory central charges.%, all of this inspired by the results in \cite{Lozano:2020sae}. 

\noindent 

\section*{Acknowledgements}

We are thankful to Niall Macpherson, Carlos Nunez, Salomon Zacarias and very especially Yolanda Lozano for useful discussions and for their comments on the draft.  

AR is supported by CONACyT-Mexico.

\appendix
%\section{1}\label{a}

\bibliography{Bibliography}

 \end{document}